\documentclass[ aps, pra, amsmath, amssymb, amsfonts, twocolumn,superscriptaddress, footinbib, longbibliography]{revtex4-1}
\usepackage{graphicx}
\usepackage{dcolumn}
\usepackage{bm}
\usepackage{amsmath}
\usepackage{amssymb}
\usepackage{hyperref} 
\usepackage{cleveref,braket,xcolor,subcaption}
\usepackage{dcolumn}
\usepackage{color}
\usepackage{mathrsfs}
\usepackage{wasysym}

\usepackage{caption}

\DeclareMathOperator{\tr}{tr}

\begin{document}

\preprint{APS/123-QED}

\title{All-optical Neural Network Quantum State Tomography}

\author{Ying Zuo} \thanks{These authors contributed equally to this work.}
\affiliation{Department of Physics, The Hong Kong University of Science and Technology, Clear Water Bay, Kowloon, Hong Kong, China}

\author{Chenfeng Cao}\thanks{These authors contributed equally to this work.}
\affiliation{Department of Physics, The Hong Kong University of Science and Technology, Clear Water Bay, Kowloon, Hong Kong, China}

\author{Ningping Cao}
\affiliation{Department of Mathematics \& Statistics, University of Guelph, Guelph N1G 2W1 ON, Canada}
\affiliation{Institute for Quantum Computing, University of Waterloo, Waterloo, ON N2L 3G1, Canada}

\author{Xuanying Lai}
\affiliation{Department of Physics, The University of Texas at Dallas, Richardson, Texas 75080, USA}

\author{Bei Zeng}%
 \email{zengb@ust.hk}
\affiliation{Department of Physics, The Hong Kong University of Science and Technology, Clear Water Bay, Kowloon, Hong Kong, China}

\author{Shengwang Du}%
 \email{dusw@utdallas.edu}
\affiliation{Department of Physics, The University of Texas at Dallas, Richardson, Texas 75080, USA}

\date{\today}

\begin{abstract}
Quantum state tomography (QST) is a crucial ingredient for almost all aspects of experimental quantum information processing. As an analog of ``imaging" technique in quantum settings, QST is born to be a data science problem, where machine learning techniques, noticeably neural networks, have been applied extensively. Here, we build and demonstrate an all-optical neural network (AONN) for photonic polarization qubit QST. The AONN is equipped with built-in optical nonlinear activation functions based on electromagnetically induced transparency. The experimental results show that our AONN can detrmine the phase parameter of the qubit state accurately. As optics is highly desired for quantum interconnections, our AONN-QST may contribute to the realization of all-optical quantum networks and inspire the ideas combining artificial optical intelligence with quantum information studies. 
\end{abstract}

\maketitle

\section{Introduction}

Quantum state tomography (QST) is a standard process of reconstructing quantum information of an unknown quantum state through measurements on its copies. QST is used to verify state preparation, exam state properties such as correlations, and calibrate experimental systems. It is a crucial part for almost all aspects of experimental quantum information processing, including quantum computing, quantum metrology, and quantum communication~\cite{bouchard2019quantum,d2003quantum,leonhardt1995quantum,thew2002qudit,lvovsky2009continuous,PhysRevLett.126.100402}.

As an analog of ``imaging" technique in quantum settings, QST is born to be a data science problem. Given limited copies of an unknown state $\rho$, we can extract its information via QST. QST is essentially an inverse problem, and such information recovering tasks are well suited to machine leaning. Quantum learning theory indicates that $\Theta\left(2^{2 n} / \varepsilon^{2}\right)$ copies of $\rho$ are necessary and sufficient to learn $\rho$ up to trace distance $\epsilon$~\cite{o2016efficient}. Although the tremendous resource requirement makes full-state QST impractical for large-scale systems, several weaker quantum learning models (e.g., probably approximately correct learning ~\cite{aaronson2007learnability}, online learning~\cite{aaronson2019online}, shadow tomography~\cite{aaronson2019shadow, huang2020predicting}) can exponentially reduce the computational resource for learning some 2-outcome measurement expectation values or ``shadows".

Artificial neural network (NN), as a powerful algorithm in machine learning to fit a specific function, has been widely used for solving quantum information problems, such as quantum optimal control~\cite{niu2019universal, an2020high}, quantum maximum entropy estimation~\cite{npcao2020supervised}, Hamiltonian reconstruction~\cite{cao2020supervised}, etc. NNs have aalso been widely applied for QST applications, such as recovering the information of local-Hamiltonian ground states from local measurements~\cite{xin2019local} efficiently, performing tomography on highly entangled state with large system size~\cite{torlai2018neural}, mitigating the state-preparation-and-measurement (SPAM) errors in experiments~\cite{palmieri2020experimental},  and improving the state fidelity~\cite{quek2018adaptive, ahmed2020classification}. Generative models with neural networks can also perform QST with dramatically lower cost~\cite{carrasquilla2019reconstructing, ahmed2020quantum}.

\begin{figure}[ht]
\centering
\includegraphics[width = \linewidth]{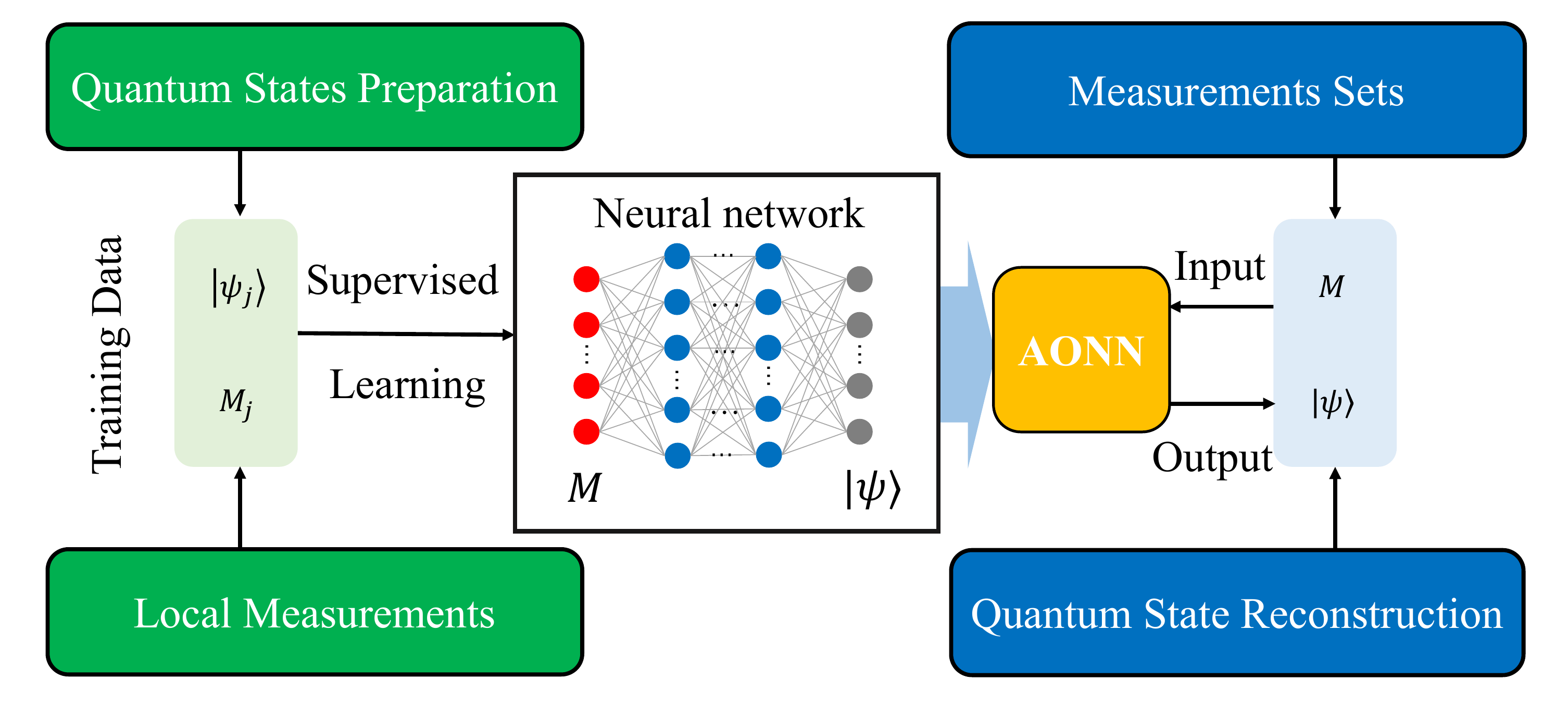}
\caption{Schematics of all-optical neural network based quantum state tomography.}
\label{fig:intro}
\end{figure}

In this work, we demonstrate neural network QST with an all-optical neural network (AONN). Several optical implementations for realizing full connected neural network hardware have been proposed and demonstrated recently\cite{wetzstein2020inference,shastri2021photonics,woods2012photonic,shen2017deep,lin2018all,zuo2019all}. Optical computing takes advantages of the bosonic wave nature of light: superposition and interference give rise to its intrinsic parallel computing ability. Meanwhile light is the fastest information carrier in nature. AONN is promising for next-generation artificial intelligence hardware which provides high energy efficiency, low crosstalk, light-speed processing, and massive parallelism. As compared to the electronic version, AONNs are ideal to deal with visual signals and information which are naturally generated and coded in light, such as image recognition and vehicular automation. However, most AONN demonstrations are still restricted to linear computation only due to the lack of suitable nonlinearity at low light level for large amount of optical neurons \cite{woods2012photonic,shen2017deep,lin2018all} . Without nonlinear activation functions, an AONN is always equivalent to a single-layer structure which cannot be applied for ``real" deep machine learning. This problem has not been solved until most recently optical nonlinearity based on electromagnetically induced transparency (EIT) \cite{zuo2019all, zuo2021scalability}, phase-change materials \cite{feldmann2019all}, and saturated absorption \cite{Guo:21,ryou2021freespace} was implemented to realize artificial optical neurons for AONNs.

Figure ~\ref{fig:intro} illustrates a general scheme of AONN QST. Firstly we collect the training data set from a known quantum state \{$|\psi_j \rangle$\} and the corresponding local measurements \{$M_j$\}. Secondly we train neural network under supervised learning with some nonlinear activation functions in its hidden neurons to obtain the optimal network parameters. Third We take the trained network parameters to configure the AONN and perform some fine adjustment to optimize the hardware performance. At last, we feed measurement data sets to the trained AONN to reconstruct unknown quantum states. To validate this scheme, in the following sections we start from a general discussion of QST with the computer simulated NN and then describe our AONN experimental approach.

\section{NN for QST}

\begin{figure}[h!]

\centering
\includegraphics[width = 0.8 \linewidth]{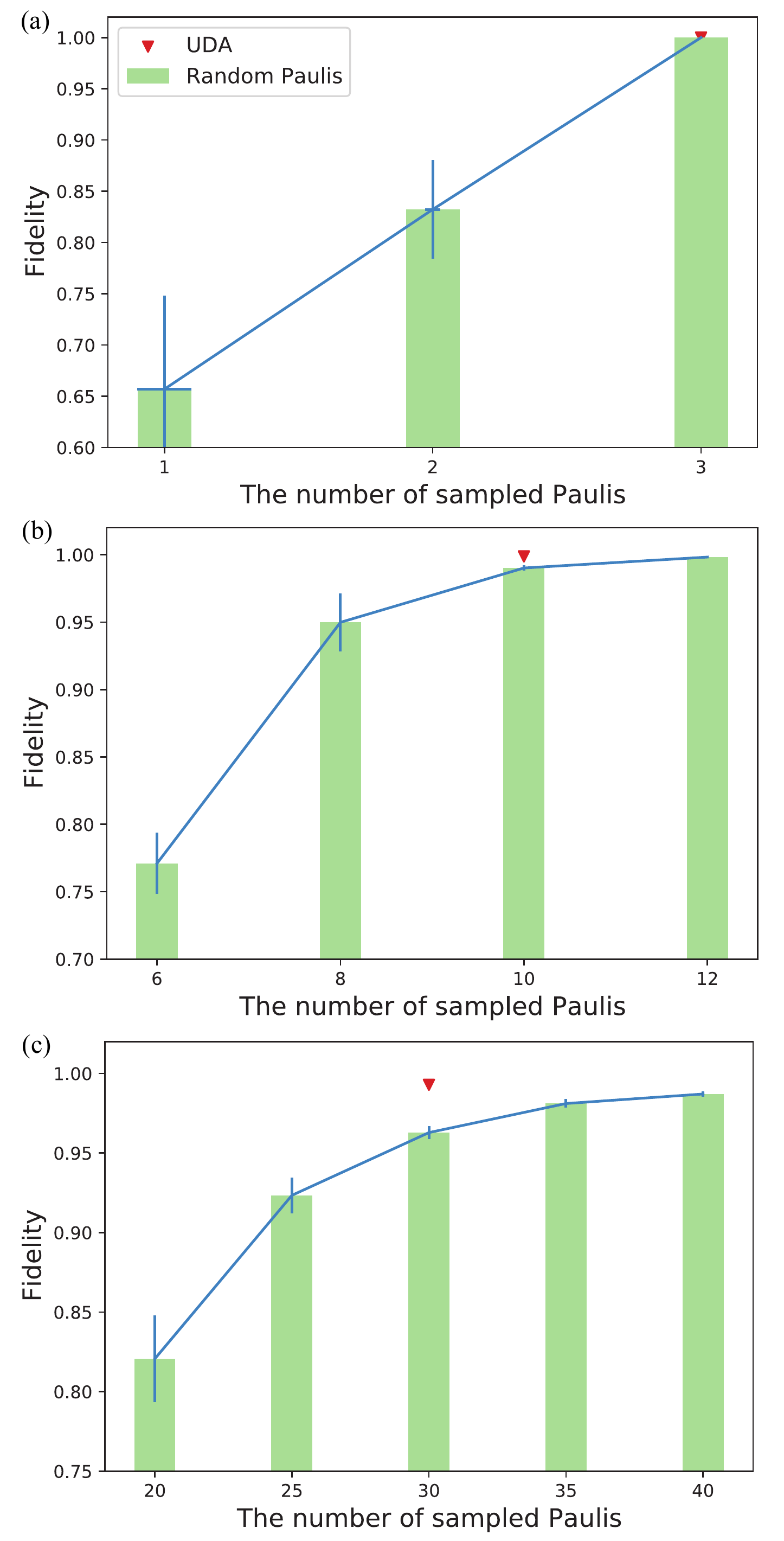}\label{fig:2q}

\caption{\textbf{The fidelities of NN predictions for different samples of Pauli operators:} The red triangles are the average fidelities for UDA Pauli operator sets, which is very close to 1. The green bars are the average fidelities for random sampled Pauli operator sets. The blue lines are the error-bars for different samples. We train NN to predict state wavefunctions from measurements for (a) 1 qubit, (b) 2 qubits and (c) 3 qubits.  }\label{fig:ONN-sim}
\end{figure}

\begin{figure*}[ht]
\centering
\includegraphics[width=\textwidth]{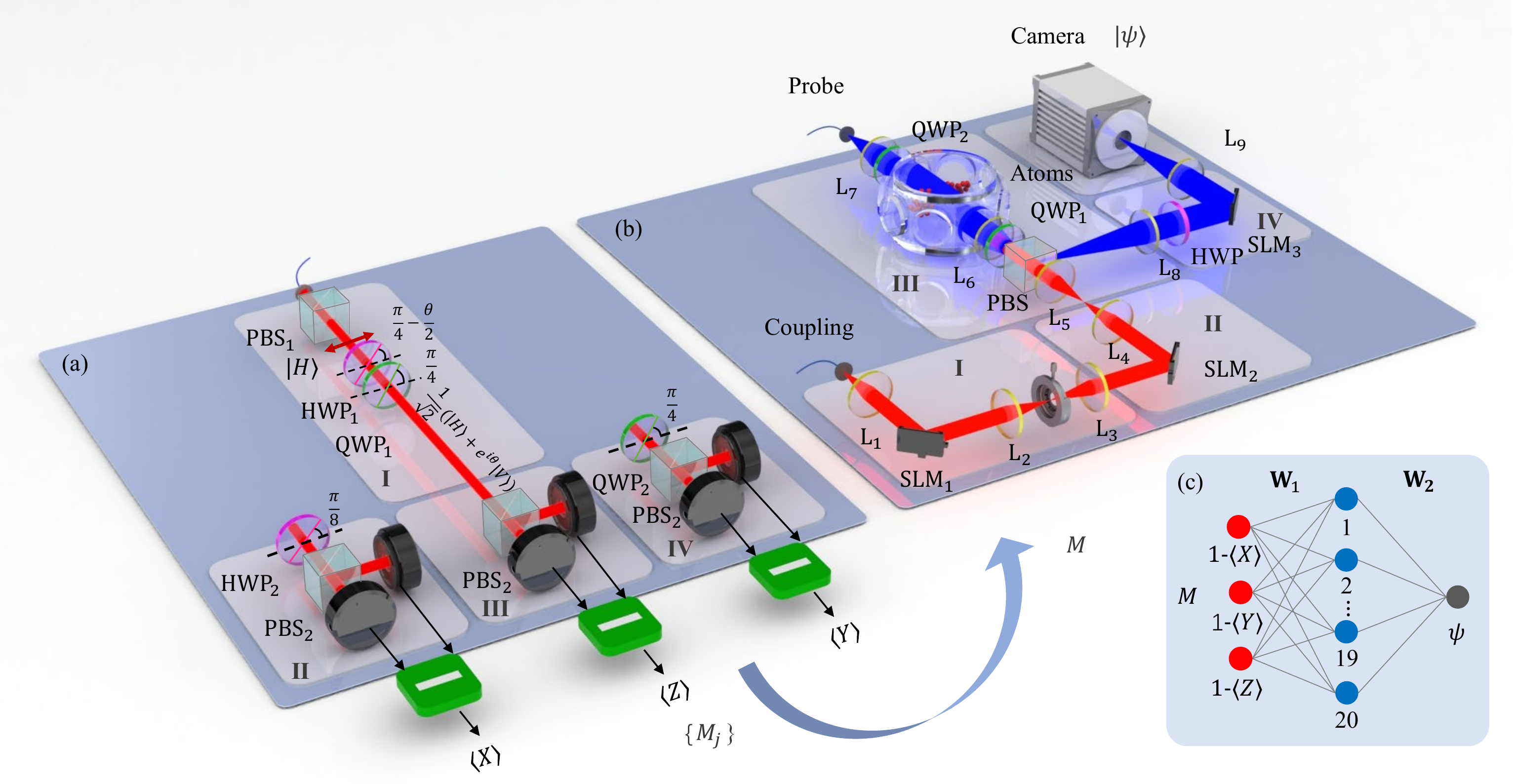}
\caption{\label{fig:system} Schematics of all-optical implementation of quantum state tomography. (a) Schematics of optical neural network. I. Input generation. II. Linear operation of first layer. III. Nonlinear operation. IV. Linear operation of second layer. Spatial light modulators: SLM1 (HOLOEY LETO), SLM2(HOLOEYE PLUTO-2), and SLM3(HOLOEY GEAE-2). Camera: Hamamatsu C11440-22CU. PBS: polarization beam splitter. Lenses: L1 ($f=100$ mm), L2 ($f=200$ mm), L3 ($f=250$ mm), L4 ($f=350$ mm), L5 ($f=350$ mm), L6 ($f=50$ mm), L7 ($f=50$ mm), L8 ($f=450$ mm) and L9 ($f=450$ mm). (b) Optical layout of qubit quantum state tomography, including generation of polarization state (top panel), measurement of $\langle Z\rangle$, $\langle X \rangle$ and $\langle Y \rangle$ (bottom panel). (c) The neural network structure employed.} 
\end{figure*}{}

We here consider a general $n$-qubit space with Pauli operators (removed the all identity term) defined as
\begin{equation}\label{eq:PaulOperators}
P = \{ \sigma_{i_1}^{(1)}\otimes \cdots\otimes \sigma^{(n)}_{i_n}| \sigma^{(k)}_{i_k} \in \mathcal{P}, \sum_{k = 1}^n i_k \neq 0  \},
\end{equation}
where $\mathcal{P} = \{\sigma_0 = I, \sigma_1 = X, \sigma_2=Y, \sigma_3=Z\}$. Every term in $P$ is specified by its index $(i_1, i_2, \cdots, i_n)$.
Measuring every element in $P$ performs a QST for any $n$-qubit quantum state $\rho$. For instance, when $n=1$,
we need to measure all three Paulis $X,Y,Z$ for QST.
Clearly, the cardinality of $P$ grows exponentially with $n$.
When $\rho$ is a pure state, one may use techniques to reduce the number of measurements for $n>1$. 
Compressed sensing is one the standard techniques to recover low-rank quantum states from randomly sampled Pauli operators~\cite{gross2010quantum,flammia2012quantum}.

When $\rho$ is a pure state, it can be written as a ket
\begin{equation}\label{eq:ket}
\ket{\psi} = \sum_{k=1}^{2^n} a_k \ket{\phi_k},
\end{equation}
where \{$\ket{\phi_k}$\} are the computational basis, and the amplitudes $a_k\in \mathbb{C}$ are normalized (i.e. $\sum_{k=1}^{2^n} (a_{k,\text{r}}^2 + a_{k,\text{im}}^2) = 1$, where $a_{k,\text{r}}\in \mathbb{R}$ and $a_{k,\text{im}}\in \mathbb{R}$ are the real and imaginary parts of $a_k$ respectively).
The measurement expectation values  $P$ are $\vec c = \tr(\rho \cdot P) = (\tr(\rho P_1),\tr(\rho P_2),\cdots,\tr(\rho P_{4^n-1}))$. 
For a single qubit pure state $\alpha|0\rangle+\beta|1\rangle$, its density matrix can be expressed as
\begin{equation}
\rho=\frac{1}{2}(1+\vec c\cdot \vec \sigma)
\end{equation} 
where $\vec \sigma = \{X,Y,Z\}$, and $\vec c = (\tr(\rho X),\tr(\rho Y),\tr(\rho Z))\equiv (\langle X\rangle, \langle Y\rangle, \langle Z\rangle)$.

For $n>1$, we consider compressed sensing to reduce the number of measured operators.
To perform compressed sensing, one needs to randomly sample a set $P^m = \{P_1,\cdots, P_m\}$ of $m$ Pauli operators from $P$, then use $\vec c = \tr(\rho \cdot P^m) = (\tr(\rho P_1),\tr(\rho P_2),\cdots,\tr(\rho P_m))$ to recover the unknown state $\rho$, more precisely, the parameters of $\rho$. This can be took as a regression problem to estimate the function between $\vec c$ and parameters of $\rho$ (e.g. $a_{k,\text{r}}$ and $a_{k,\text{im}}$).

NNs are excellent tools for solving regression problems. In the compressed sensing inspired QST, the expectation values $\vec c$ from random-sampled $P^m$ are inputs to the network, and ($a_{k,\text{r}}$, $a_{k,\text{im}}$) are the outputs. Note that there are different types of NNs with various structures and training procedures. According to the law of parsimony (Occam's razor), we use the simplest type of NNs in this letter -- fully-connected, feed-forward NNs. That is, the neurons between nearest layers are fully connected with one another and the information only pass forward while training. The supervised training process is to compare the ideal outputs $(a_{k,\text{r}}, a_{k,\text{im}})$ with current NN outputs, and update parameters embedded in the NN to minimize their difference.

We numerically trained computer-based NNs nonlinear activation functions for 1-qubit, 2-qubit and 3-qubit QST. For the 1-qubit system, the number of sampled operators $m \in [1,2,3]$; for the 2-qubit system, the number of sampled operators $m \in [6,8,10,12]$; for the 3-qubit system, $m \in [20,25,30,35,40]$. Plainly, $m$ equals the number of input neurons, and $n$ decides the number of output neurons. For each $m$, three sets of Pauli operators have been sampled and tested. Figure \ref{fig:ONN-sim} plots the average fidelities (green bars) of both cases as functions of the number of randomly sampled Paulis. For the single qubit system, the fidelity reaches 99.99\%  with 3 paulis. For the 2-qubit system, the fidelity reaches 99.9\% with 10 randomly sampled paulis [Fig.~\ref{fig:ONN-sim}(a)]. For the 3-qubit system, a fidelity of higher than 99.9\% requires more than 35 randomly sampled Paulis  [Fig.~\ref{fig:ONN-sim}(b)]. Details of training can be found in Appendix \ref{app:ONN-sim}.

Theoretically, a pure state $\rho$ is Uniquely Determined among All states (UDA) of a set of operators $F$ means that there is no other state, pure or mix, has the same expectation values while measuring $F$~\cite{chen2013uniqueness}.  In Ref.~\cite{ma2016pure}, authors discovered two sets of Pauli operators, $P_\text{2-UDA}$ and $P_\text{3-UDA}$, that are UDA for all 2-qubit and 3-qubit pure states respectively. (See Appendix \ref{app:UDA-op} for the particular sets $P_\text{2-UDA}$ and $P_\text{3-UDA}$.) Namely, they are special cases of Pauli operator sets that the map between expectation values and the measured state $\rho$ is bijective. Similarly, we apply NNs for these two sets of UDA operators and obtain the prediction fidelities of 99.9\% for the 2-qubit case and 99.3\% for the 3-qubit case (red triangles in Figure \ref{fig:ONN-sim}).

We remark that our UDA scheme is not readily scalable for larger systems, however there exist protocals with better scalability, e.g., compressed sensing~\cite{gross2010quantum}, shadow tomography~\cite{aaronson2019shadow, huang2020predicting}, where NN can be also be naturally used to enhance the protocols. Also, our NN-based scheme can be adapted to quantum tomography in the optical system, by taking into account of physical constrains, which will be discussed in detailed in the next section.

\section{AONN-QST experiment}\label{sec:3}

In this first proof-of-principle experimental demonstration, we implement the single qubit space with light polarizations, i.e., horizontal polarization $|H\rangle=|0\rangle$ and vertical polarization $|V\rangle=|1\rangle$. Instead of making a full QST, here we focus our task to determine the phase parameter of a pure state $|\psi\rangle=\frac{1}{\sqrt{2}}(|H\rangle+e^{i\theta}|V\rangle)$. The experimental AONN-QST setup is displayed in Fig. ~\ref{fig:system}. In conventional QST, an arbitrary polarization state can be reconstructed by measuring the expectation values of the three Pauli operators. Figure \ref{fig:system}(a) illustrates such an optical measurement setup. A laser beam passes through a polarization beam splitter (PBS$_1$) and becomes horizontally polarized ($|H\rangle$). The target state $|\psi\rangle=\frac{1}{\sqrt{2}}(|H\rangle+e^{i\theta}|V\rangle)$ is prepared by passing this horizontally polarized light through a half-wave plate (HWP$_1$) and a quarter-wave plate (QWP$_1$). The fast axis of the HWP$_1$ is aligned with an angle $\pi/4-\theta/2$ to the horizontal direction. The fast axis of the QWP$_1$ is aligned with an angle $\pi/4$ to the horizontal direction. $\langle X \rangle$, $\langle Y \rangle $ and $\langle Z\rangle$ are obtained by sending the light polarization qubit state to the measurement units II, III and IV shown in Fig.~\ref{fig:system}(a). To determine $\langle Z \rangle$, we send the polarization qubit directly to PBS$_2$ which projects $|H\rangle$ and $|V\rangle$ into two photodetectors in the measurement unit III. The normalized differential output from these two photodetectors gives the value $\langle Z \rangle$. The same setup can also be used to determine $\langle X \rangle$ or $\langle Y \rangle$ by placing HWP$_2$ or QWP$_2$ before PBS$_2$ as shown in II or IV, respectively (see Appendix~\ref{app:polar} for details).

We obtain a data set \{$M_i$\}=$\{|\phi_i\rangle: 1-\langle X \rangle_i,1-\langle Y \rangle_i,1-\langle Z \rangle_i\}$ by varying the phase $\theta \in [0, \pi/2]$ in the qubit state $|\psi\rangle=\frac{1}{\sqrt{2}}(|H\rangle+e^{i\theta}|V\rangle)$ and use them to train our AONN in Fig.~\ref{fig:system}(b). The AONN comprises an input layer of 3 neurons, a hidden layer of 20 neurons and a single-neuron output layer \cite{zuo2019all,zuo2021scalability}.  Figure \ref{fig:system}(b) shows the optical layout of the AONN and its network structure diagram is displayed in Fig.~\ref{fig:system}(c). The three coupling laser beams in the optical input layer are generated by a spatial light modulator (SLM1) [Fig.~\ref{fig:system}(b)], lenses L2 and L3, and an aperture, as shown in unit I of Fig.\ref{fig:system}(b).
The SLM1 is divided into 3 parts and each part is encoded with sine phase pattern $m \pi\sin(\frac{2\pi}{T_{mi}}i+\frac{2\pi}{T_{mj}}j)$, where $m$ is the modulation depth, $T_{mi}$ and $T_{mj}$ are the period of modulation along $x$ and $y$ directions and $i$ and $j$ are the pixel number along $x$ and $y$ directions. The sine phase encoded on SLM1 modulate the beams into separated beams at focal plane of lens L2 and the aperture behaves as a filter to keep the zero-order beam, whose intensity is determined by the modulation depth $m$. Thus, the three beams with designed intensity in generated and collimated propagate to the SLM2 through lens L3. These weighted beams, as the input vector, are incident on SLM2 which diffracts each beam into 20 directions with designed weight (See Appendix \ref{app:gsw} for the algorithm to calculate the pattern encoded on SLM2). A Fourier lens L4 performs linear summation for the beams diffracted into the same direction and forms 20 spots on its front focal plane. Thus, the combination of SLM2 and L4 completes the first linear operation W$_1$ and generates the input to the hidden layer. We then image these 20 spots with lenses L5 and L6 to laser-cooled $^{85}$Rb atoms in a two-dimensional (2D) magneto-optical trap (MOT) \cite{metcalf2003laser, 2DMOTRSI2012}, where these 20-spot coupling beam pattern spatially modulate the transparency of the atomic medium through electromagnetically induced transparency (EIT) \cite{EITHarris,fleischhauer2005electromagnetically}. Another relatively weak collimated probe beam counter-propagates through the MOT and its spatial transmission is nonlinearly controlled by the 20-spot coupling beam pattern. Here the nonlinear optical activation functions are realized with EIT in cold atoms. The equation of nonlinear activation functions follows Equation \ref{eqn:EIT}. The image of the probe beam transmission pattern by lens L6 and L8 becomes the output of the 20 hidden neurons. SLM3 and Fourier lens L9 perform the second linear matrix operation W$_2$ and the output is recorded by a camera. The technical details of our AONN is described in ref. \cite{zuo2019all,zuo2021scalability}. 

In this work because we encode data into light energy, the AONN can only handle positive values: Input, output, linear matrix elements, and input/output of nonlinear activation functions are all positive values \cite{zuo2019all,zuo2021scalability}. Meanwhile, the EIT optical nonlinear activation functions are increasing and convex. Therefore the AONN is only able to perform regression task on increasing and convex functions. To match the AONN constrains, we performing a transform to the input variable, e.g., $\langle X \rangle$ to 1-$\langle X \rangle$, so that all input values to the AONN nodes are positive. We add these conditions to NN to simulate the AONN performance. We find that this specific AONN fails to describe the whole range of nonmonotonic functions. For the first proof-of-principle experimental demonstration, we will only apply the AONN for single-qubit QST with phase $\theta$ within $[0,\pi/2]$. It is surprising that such a positive-valued AONN is still able to perform some types of QST.

To train the AONN, we prepared the training data set \{$M_i$\} from 23 phase values from a uniform distribution $\theta_{j} \sim U(0,\pi/2)$, corresponding to the optical polarization states $\{\rho_{j} = \mathcal{N}(|\phi_{j}\rangle\langle\phi_{j}|)\}$. Here $\mathcal{N}$ is the noise channel in experiments, and measure the Pauli expectation values $\langle X \rangle$, $\langle Y \rangle$, $\langle Z \rangle$. In a similar way, we prepare a test set with 32 independent data samples. 

In addition to optical quantum states, we sample data from the IBM quantum computer ibmq\_ourense~\cite{ibmq_5_ourense}, and implement the same AONN training for comparison. The quantum circuit to prepare $|\psi\rangle = (|H\rangle + e^{i\theta}|V\rangle)/\sqrt{2}$ is the initial state $|H\rangle$ going through a Hadamard gate and then going through a RZ rotation gate. On ibmq\_ourense, we uniformly sample 158 data points as training set, 50 data points as test set. Both experimental optical quantum state tomography and IBMQ tomography data are used to train two NNs. Details of training AONN can be found in Appendix \ref{app:AONN-train}.

\begin{figure}
\includegraphics[width=\linewidth]{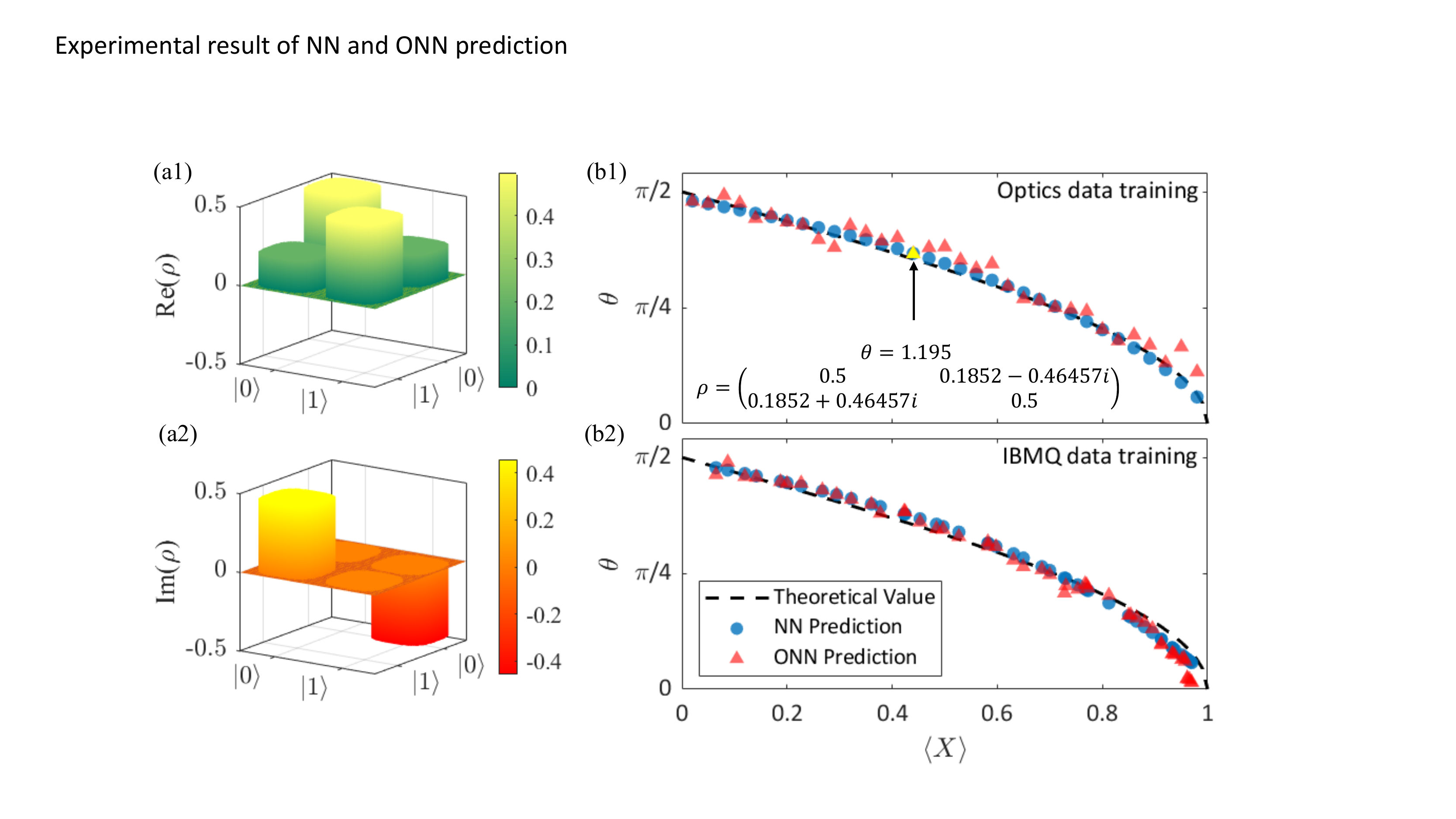}
\centering
\caption{\label{fig:res} (a) Optical tomography of qubit (b) Experimental AONN tomography result. The AONN is training by optical tomography data (b1) and IBMQ tomography data (b2). The black dashed line is the theoretical value of the phase $\theta$ according to the $\langle X \rangle$. The blue circles are the phase $\theta$ numerically predicted by the trained neural network and the red triangles are the experimentally measured predictions of $\theta$ according to $\langle X \rangle$}
\end{figure}

Figure \ref{fig:res} shows the AONN state construction results using neural network models trained by the AONN QST training set and the IBM quantum computer training set separately. The theoretical value is calculated from $\langle X \rangle$ directly. With the AONN system set up to the training results, we sent a set of input vector to the system. The example of real and imaginary part of density matrix are shown in Fig.~\ref{fig:res}(a). The experimentally measured state example is predicted by AONN QST training model. The example input vector for AONN model is $(\langle X\rangle, \langle Y \rangle, \langle Z \rangle )= (0.440, 0.898, 0)$ and the AONN experimental predicted state is $\theta=1.195$ and
 $\rho =
 \begin{pmatrix} 
 0.5 & 0.1852-0.46457i \\ 
 0.1852+0.46457i & 0.5
 \end{pmatrix}
 $ which is close to the theoretical value $\theta=1.1152$ and neural network predicted value $\theta = 1.1532$. The state is also marked with yellow triangle in Fig.~\ref{fig:res} (b1). The experimental results are shown in Fig.~\ref{fig:res}(b). The theoretical value, NN predicted value and experimental AONN predicted value collapse together for both optics data training (Fig.~\ref{fig:res} (b1) shows) and IBMQ data training (Fig.~\ref{fig:res} (b2) shows). The theoretical value, neural network prediction value and AONN predicted value are consistent in both cases. The results suggest that our positive-valued AONN with EIT nonlinear activation functions is capable for qubit QST.

\section{Discussion and conclusion}

While most demonstrations of optical neural networks took classification tasks to verify their feasibility\cite{lin2018all,feldmann2019all,ShenNP2017}, we performed the first AONN-QST. To accomplish regression tasks, nonlinear function is essential as long as the relation between input vector and output vector cannot be expressed linearly. The tunable EIT nonlinear optical activation functions in our AONN offer opportunities for performing regression tasks with certain functions. Although our AONN has some certain limitation that the linear operation matrix elements are all positive valued, it has potential to solve some QST problems as we have demonstrated here for single qubit tomography. We believe the next generation of complex-valued AONNs with encoding data in both light amplitude and phase will be more powerful. The future development of complex-valued AONNs may become a more powerful tool for full QST in a much larger $n$-qubit space.

Optical quantum network~\cite{Kimble2008} has been brought to the fore by the reduced decoherence and high speed of photons. Recently, apart from generating optical quantum states~\cite{Gu2019} and optical quantum communication over long distance~\cite{Yu2020}, multiple state-of-the-art experiments on optical quantum interfaces to store~\cite{Li2020} and distribute entanglements~\cite{Choi2010,Pu2017} have been exhibited. Among all of these, QST is essential for not only characterizing generation and preservation of quantum states but also has the potential to verifying the entanglement distributed across the whole network. We believe that our all-optical setup of integrated AONN-QST will shed light on replenishing the all-optical quantum network with one more brick. 
We acknowledge the use of IBM Quantum services for this work.

\bibliography{ONN.bib}

\begin{thebibliography}{51}%
\makeatletter
\providecommand \@ifxundefined [1]{%
 \@ifx{#1\undefined}
}%
\providecommand \@ifnum [1]{%
 \ifnum #1\expandafter \@firstoftwo
 \else \expandafter \@secondoftwo
 \fi
}%
\providecommand \@ifx [1]{%
 \ifx #1\expandafter \@firstoftwo
 \else \expandafter \@secondoftwo
 \fi
}%
\providecommand \natexlab [1]{#1}%
\providecommand \enquote  [1]{``#1''}%
\providecommand \bibnamefont  [1]{#1}%
\providecommand \bibfnamefont [1]{#1}%
\providecommand \citenamefont [1]{#1}%
\providecommand \href@noop [0]{\@secondoftwo}%
\providecommand \href [0]{\begingroup \@sanitize@url \@href}%
\providecommand \@href[1]{\@@startlink{#1}\@@href}%
\providecommand \@@href[1]{\endgroup#1\@@endlink}%
\providecommand \@sanitize@url [0]{\catcode `\\12\catcode `\$12\catcode
  `\&12\catcode `\#12\catcode `\^12\catcode `\_12\catcode `\%12\relax}%
\providecommand \@@startlink[1]{}%
\providecommand \@@endlink[0]{}%
\providecommand \url  [0]{\begingroup\@sanitize@url \@url }%
\providecommand \@url [1]{\endgroup\@href {#1}{\urlprefix }}%
\providecommand \urlprefix  [0]{URL }%
\providecommand \Eprint [0]{\href }%
\providecommand \doibase [0]{http://dx.doi.org/}%
\providecommand \selectlanguage [0]{\@gobble}%
\providecommand \bibinfo  [0]{\@secondoftwo}%
\providecommand \bibfield  [0]{\@secondoftwo}%
\providecommand \translation [1]{[#1]}%
\providecommand \BibitemOpen [0]{}%
\providecommand \bibitemStop [0]{}%
\providecommand \bibitemNoStop [0]{.\EOS\space}%
\providecommand \EOS [0]{\spacefactor3000\relax}%
\providecommand \BibitemShut  [1]{\csname bibitem#1\endcsname}%
\let\auto@bib@innerbib\@empty
\bibitem [{\citenamefont {Bouchard}\ \emph {et~al.}(2019)\citenamefont
  {Bouchard}, \citenamefont {Hufnagel}, \citenamefont {Koutn{\`y}},
  \citenamefont {Abbas}, \citenamefont {Sit}, \citenamefont {Heshami},
  \citenamefont {Fickler},\ and\ \citenamefont {Karimi}}]{bouchard2019quantum}%
  \BibitemOpen
  \bibfield  {author} {\bibinfo {author} {\bibfnamefont {Fr{\'e}d{\'e}ric}\
  \bibnamefont {Bouchard}}, \bibinfo {author} {\bibfnamefont {Felix}\
  \bibnamefont {Hufnagel}}, \bibinfo {author} {\bibfnamefont {Dominik}\
  \bibnamefont {Koutn{\`y}}}, \bibinfo {author} {\bibfnamefont {Aazad}\
  \bibnamefont {Abbas}}, \bibinfo {author} {\bibfnamefont {Alicia}\
  \bibnamefont {Sit}}, \bibinfo {author} {\bibfnamefont {Khabat}\ \bibnamefont
  {Heshami}}, \bibinfo {author} {\bibfnamefont {Robert}\ \bibnamefont
  {Fickler}}, \ and\ \bibinfo {author} {\bibfnamefont {Ebrahim}\ \bibnamefont
  {Karimi}},\ }\bibfield  {title} {\enquote {\bibinfo {title} {Quantum process
  tomography of a high-dimensional quantum communication channel},}\
  }\href@noop {} {\bibfield  {journal} {\bibinfo  {journal} {Quantum}\ }\textbf
  {\bibinfo {volume} {3}},\ \bibinfo {pages} {138} (\bibinfo {year}
  {2019})}\BibitemShut {NoStop}%
\bibitem [{\citenamefont {D'Ariano}\ \emph {et~al.}(2003)\citenamefont
  {D'Ariano}, \citenamefont {Paris},\ and\ \citenamefont
  {Sacchi}}]{d2003quantum}%
  \BibitemOpen
  \bibfield  {author} {\bibinfo {author} {\bibfnamefont {G~Mauro}\ \bibnamefont
  {D'Ariano}}, \bibinfo {author} {\bibfnamefont {Matteo~GA}\ \bibnamefont
  {Paris}}, \ and\ \bibinfo {author} {\bibfnamefont {Massimiliano~F}\
  \bibnamefont {Sacchi}},\ }\bibfield  {title} {\enquote {\bibinfo {title}
  {Quantum tomography},}\ }\href@noop {} {\bibfield  {journal} {\bibinfo
  {journal} {Advances in Imaging and Electron Physics}\ }\textbf {\bibinfo
  {volume} {128}},\ \bibinfo {pages} {206--309} (\bibinfo {year}
  {2003})}\BibitemShut {NoStop}%
\bibitem [{\citenamefont {Leonhardt}(1995)}]{leonhardt1995quantum}%
  \BibitemOpen
  \bibfield  {author} {\bibinfo {author} {\bibfnamefont {Ulf}\ \bibnamefont
  {Leonhardt}},\ }\bibfield  {title} {\enquote {\bibinfo {title} {Quantum-state
  tomography and discrete wigner function},}\ }\href@noop {} {\bibfield
  {journal} {\bibinfo  {journal} {Phys. Rev. Lett.}\ }\textbf {\bibinfo
  {volume} {74}},\ \bibinfo {pages} {4101} (\bibinfo {year}
  {1995})}\BibitemShut {NoStop}%
\bibitem [{\citenamefont {Thew}\ \emph {et~al.}(2002)\citenamefont {Thew},
  \citenamefont {Nemoto}, \citenamefont {White},\ and\ \citenamefont
  {Munro}}]{thew2002qudit}%
  \BibitemOpen
  \bibfield  {author} {\bibinfo {author} {\bibfnamefont {RT}~\bibnamefont
  {Thew}}, \bibinfo {author} {\bibfnamefont {Kae}\ \bibnamefont {Nemoto}},
  \bibinfo {author} {\bibfnamefont {Andrew~G}\ \bibnamefont {White}}, \ and\
  \bibinfo {author} {\bibfnamefont {William~J}\ \bibnamefont {Munro}},\
  }\bibfield  {title} {\enquote {\bibinfo {title} {Qudit quantum-state
  tomography},}\ }\href@noop {} {\bibfield  {journal} {\bibinfo  {journal}
  {Phys. Rev. A}\ }\textbf {\bibinfo {volume} {66}},\ \bibinfo {pages} {012303}
  (\bibinfo {year} {2002})}\BibitemShut {NoStop}%
\bibitem [{\citenamefont {Lvovsky}\ and\ \citenamefont
  {Raymer}(2009)}]{lvovsky2009continuous}%
  \BibitemOpen
  \bibfield  {author} {\bibinfo {author} {\bibfnamefont {Alexander~I}\
  \bibnamefont {Lvovsky}}\ and\ \bibinfo {author} {\bibfnamefont {Michael~G}\
  \bibnamefont {Raymer}},\ }\bibfield  {title} {\enquote {\bibinfo {title}
  {Continuous-variable optical quantum-state tomography},}\ }\href@noop {}
  {\bibfield  {journal} {\bibinfo  {journal} {Rev. Mod. Phys.}\ }\textbf
  {\bibinfo {volume} {81}},\ \bibinfo {pages} {299} (\bibinfo {year}
  {2009})}\BibitemShut {NoStop}%
\bibitem [{\citenamefont {Rambach}\ \emph {et~al.}(2021)\citenamefont
  {Rambach}, \citenamefont {Qaryan}, \citenamefont {Kewming}, \citenamefont
  {Ferrie}, \citenamefont {White},\ and\ \citenamefont
  {Romero}}]{PhysRevLett.126.100402}%
  \BibitemOpen
  \bibfield  {author} {\bibinfo {author} {\bibfnamefont {Markus}\ \bibnamefont
  {Rambach}}, \bibinfo {author} {\bibfnamefont {Mahdi}\ \bibnamefont {Qaryan}},
  \bibinfo {author} {\bibfnamefont {Michael}\ \bibnamefont {Kewming}}, \bibinfo
  {author} {\bibfnamefont {Christopher}\ \bibnamefont {Ferrie}}, \bibinfo
  {author} {\bibfnamefont {Andrew~G.}\ \bibnamefont {White}}, \ and\ \bibinfo
  {author} {\bibfnamefont {Jacquiline}\ \bibnamefont {Romero}},\ }\bibfield
  {title} {\enquote {\bibinfo {title} {Robust and efficient high-dimensional
  quantum state tomography},}\ }\href {\doibase 10.1103/PhysRevLett.126.100402}
  {\bibfield  {journal} {\bibinfo  {journal} {Phys. Rev. Lett.}\ }\textbf
  {\bibinfo {volume} {126}},\ \bibinfo {pages} {100402} (\bibinfo {year}
  {2021})}\BibitemShut {NoStop}%
\bibitem [{\citenamefont {O'Donnell}\ and\ \citenamefont
  {Wright}(2016)}]{o2016efficient}%
  \BibitemOpen
  \bibfield  {author} {\bibinfo {author} {\bibfnamefont {Ryan}\ \bibnamefont
  {O'Donnell}}\ and\ \bibinfo {author} {\bibfnamefont {John}\ \bibnamefont
  {Wright}},\ }\bibfield  {title} {\enquote {\bibinfo {title} {Efficient
  quantum tomography},}\ }in\ \href@noop {} {\emph {\bibinfo {booktitle}
  {Proceedings of the forty-eighth annual ACM symposium on Theory of
  Computing}}}\ (\bibinfo {year} {2016})\ pp.\ \bibinfo {pages}
  {899--912}\BibitemShut {NoStop}%
\bibitem [{\citenamefont {Aaronson}(2007)}]{aaronson2007learnability}%
  \BibitemOpen
  \bibfield  {author} {\bibinfo {author} {\bibfnamefont {Scott}\ \bibnamefont
  {Aaronson}},\ }\bibfield  {title} {\enquote {\bibinfo {title} {The
  learnability of quantum states},}\ }\href@noop {} {\bibfield  {journal}
  {\bibinfo  {journal} {Proc. R. Soc. A.}\ }\textbf {\bibinfo {volume} {463}},\
  \bibinfo {pages} {3089--3114} (\bibinfo {year} {2007})}\BibitemShut {NoStop}%
\bibitem [{\citenamefont {Aaronson}\ \emph {et~al.}(2019)\citenamefont
  {Aaronson}, \citenamefont {Chen}, \citenamefont {Hazan}, \citenamefont
  {Kale},\ and\ \citenamefont {Nayak}}]{aaronson2019online}%
  \BibitemOpen
  \bibfield  {author} {\bibinfo {author} {\bibfnamefont {Scott}\ \bibnamefont
  {Aaronson}}, \bibinfo {author} {\bibfnamefont {Xinyi}\ \bibnamefont {Chen}},
  \bibinfo {author} {\bibfnamefont {Elad}\ \bibnamefont {Hazan}}, \bibinfo
  {author} {\bibfnamefont {Satyen}\ \bibnamefont {Kale}}, \ and\ \bibinfo
  {author} {\bibfnamefont {Ashwin}\ \bibnamefont {Nayak}},\ }\bibfield  {title}
  {\enquote {\bibinfo {title} {Online learning of quantum states},}\
  }\href@noop {} {\bibfield  {journal} {\bibinfo  {journal} {J. Stat. Mech.}\
  }\textbf {\bibinfo {volume} {2019}},\ \bibinfo {pages} {124019} (\bibinfo
  {year} {2019})}\BibitemShut {NoStop}%
\bibitem [{\citenamefont {Aaronson}(2019)}]{aaronson2019shadow}%
  \BibitemOpen
  \bibfield  {author} {\bibinfo {author} {\bibfnamefont {Scott}\ \bibnamefont
  {Aaronson}},\ }\bibfield  {title} {\enquote {\bibinfo {title} {Shadow
  tomography of quantum states},}\ }\href@noop {} {\bibfield  {journal}
  {\bibinfo  {journal} {SIAM Journal on Computing}\ }\textbf {\bibinfo {volume}
  {49}},\ \bibinfo {pages} {STOC18--368} (\bibinfo {year} {2019})}\BibitemShut
  {NoStop}%
\bibitem [{\citenamefont {Huang}\ \emph {et~al.}(2020)\citenamefont {Huang},
  \citenamefont {Kueng},\ and\ \citenamefont {Preskill}}]{huang2020predicting}%
  \BibitemOpen
  \bibfield  {author} {\bibinfo {author} {\bibfnamefont {Hsin-Yuan}\
  \bibnamefont {Huang}}, \bibinfo {author} {\bibfnamefont {Richard}\
  \bibnamefont {Kueng}}, \ and\ \bibinfo {author} {\bibfnamefont {John}\
  \bibnamefont {Preskill}},\ }\bibfield  {title} {\enquote {\bibinfo {title}
  {Predicting many properties of a quantum system from very few
  measurements},}\ }\href@noop {} {\bibfield  {journal} {\bibinfo  {journal}
  {Nature Physics}\ }\textbf {\bibinfo {volume} {16}},\ \bibinfo {pages}
  {1050--1057} (\bibinfo {year} {2020})}\BibitemShut {NoStop}%
\bibitem [{\citenamefont {Niu}\ \emph {et~al.}(2019)\citenamefont {Niu},
  \citenamefont {Boixo}, \citenamefont {Smelyanskiy},\ and\ \citenamefont
  {Neven}}]{niu2019universal}%
  \BibitemOpen
  \bibfield  {author} {\bibinfo {author} {\bibfnamefont {Murphy~Yuezhen}\
  \bibnamefont {Niu}}, \bibinfo {author} {\bibfnamefont {Sergio}\ \bibnamefont
  {Boixo}}, \bibinfo {author} {\bibfnamefont {Vadim~N}\ \bibnamefont
  {Smelyanskiy}}, \ and\ \bibinfo {author} {\bibfnamefont {Hartmut}\
  \bibnamefont {Neven}},\ }\bibfield  {title} {\enquote {\bibinfo {title}
  {Universal quantum control through deep reinforcement learning},}\
  }\href@noop {} {\bibfield  {journal} {\bibinfo  {journal} {npj Quantum Inf.}\
  }\textbf {\bibinfo {volume} {5}},\ \bibinfo {pages} {1--8} (\bibinfo {year}
  {2019})}\BibitemShut {NoStop}%
\bibitem [{\citenamefont {An}\ \emph {et~al.}(2020)\citenamefont {An},
  \citenamefont {He}, \citenamefont {Song},\ and\ \citenamefont
  {Zhou}}]{an2020high}%
  \BibitemOpen
  \bibfield  {author} {\bibinfo {author} {\bibfnamefont {Zheng}\ \bibnamefont
  {An}}, \bibinfo {author} {\bibfnamefont {Qi-Kai}\ \bibnamefont {He}},
  \bibinfo {author} {\bibfnamefont {Hai-Jing}\ \bibnamefont {Song}}, \ and\
  \bibinfo {author} {\bibfnamefont {DL}~\bibnamefont {Zhou}},\ }\bibfield
  {title} {\enquote {\bibinfo {title} {High dimensional quantum optimal control
  with reinforcement learning},}\ }\href@noop {} {\bibfield  {journal}
  {\bibinfo  {journal} {arXiv:2007.00838}\ } (\bibinfo {year}
  {2020})}\BibitemShut {NoStop}%
\bibitem [{\citenamefont {Cao}\ \emph {et~al.}(2021)\citenamefont {Cao},
  \citenamefont {Xie}, \citenamefont {Zhang}, \citenamefont {Hou},
  \citenamefont {Zhang},\ and\ \citenamefont {Zeng}}]{npcao2020supervised}%
  \BibitemOpen
  \bibfield  {author} {\bibinfo {author} {\bibfnamefont {Ningping}\
  \bibnamefont {Cao}}, \bibinfo {author} {\bibfnamefont {Jie}\ \bibnamefont
  {Xie}}, \bibinfo {author} {\bibfnamefont {Aonan}\ \bibnamefont {Zhang}},
  \bibinfo {author} {\bibfnamefont {Shi-Yao}\ \bibnamefont {Hou}}, \bibinfo
  {author} {\bibfnamefont {Lijian}\ \bibnamefont {Zhang}}, \ and\ \bibinfo
  {author} {\bibfnamefont {Bei}\ \bibnamefont {Zeng}},\ }\bibfield  {title}
  {\enquote {\bibinfo {title} {Neural networks for quantum inverse problems},}\
  }\href@noop {} {\bibfield  {journal} {\bibinfo  {journal} {arXiv:2005.01540}\
  } (\bibinfo {year} {2021})}\BibitemShut {NoStop}%
\bibitem [{\citenamefont {Cao}\ \emph {et~al.}(2020)\citenamefont {Cao},
  \citenamefont {Hou}, \citenamefont {Cao},\ and\ \citenamefont
  {Zeng}}]{cao2020supervised}%
  \BibitemOpen
  \bibfield  {author} {\bibinfo {author} {\bibfnamefont {Chenfeng}\
  \bibnamefont {Cao}}, \bibinfo {author} {\bibfnamefont {Shi-Yao}\ \bibnamefont
  {Hou}}, \bibinfo {author} {\bibfnamefont {Ningping}\ \bibnamefont {Cao}}, \
  and\ \bibinfo {author} {\bibfnamefont {Bei}\ \bibnamefont {Zeng}},\
  }\bibfield  {title} {\enquote {\bibinfo {title} {Supervised learning in
  hamiltonian reconstruction from local measurements on eigenstates},}\
  }\href@noop {} {\bibfield  {journal} {\bibinfo  {journal} {J. Phys.: Condens.
  Matter}\ }\textbf {\bibinfo {volume} {33}},\ \bibinfo {pages} {064002}
  (\bibinfo {year} {2020})}\BibitemShut {NoStop}%
\bibitem [{\citenamefont {Xin}\ \emph {et~al.}(2019)\citenamefont {Xin},
  \citenamefont {Lu}, \citenamefont {Cao}, \citenamefont {Anikeeva},
  \citenamefont {Lu}, \citenamefont {Li}, \citenamefont {Long},\ and\
  \citenamefont {Zeng}}]{xin2019local}%
  \BibitemOpen
  \bibfield  {author} {\bibinfo {author} {\bibfnamefont {Tao}\ \bibnamefont
  {Xin}}, \bibinfo {author} {\bibfnamefont {Sirui}\ \bibnamefont {Lu}},
  \bibinfo {author} {\bibfnamefont {Ningping}\ \bibnamefont {Cao}}, \bibinfo
  {author} {\bibfnamefont {Galit}\ \bibnamefont {Anikeeva}}, \bibinfo {author}
  {\bibfnamefont {Dawei}\ \bibnamefont {Lu}}, \bibinfo {author} {\bibfnamefont
  {Jun}\ \bibnamefont {Li}}, \bibinfo {author} {\bibfnamefont {Guilu}\
  \bibnamefont {Long}}, \ and\ \bibinfo {author} {\bibfnamefont {Bei}\
  \bibnamefont {Zeng}},\ }\bibfield  {title} {\enquote {\bibinfo {title}
  {Local-measurement-based quantum state tomography via neural networks},}\
  }\href@noop {} {\bibfield  {journal} {\bibinfo  {journal} {npj Quantum Inf.}\
  }\textbf {\bibinfo {volume} {5}},\ \bibinfo {pages} {1--8} (\bibinfo {year}
  {2019})}\BibitemShut {NoStop}%
\bibitem [{\citenamefont {Torlai}\ \emph {et~al.}(2018)\citenamefont {Torlai},
  \citenamefont {Mazzola}, \citenamefont {Carrasquilla}, \citenamefont
  {Troyer}, \citenamefont {Melko},\ and\ \citenamefont
  {Carleo}}]{torlai2018neural}%
  \BibitemOpen
  \bibfield  {author} {\bibinfo {author} {\bibfnamefont {Giacomo}\ \bibnamefont
  {Torlai}}, \bibinfo {author} {\bibfnamefont {Guglielmo}\ \bibnamefont
  {Mazzola}}, \bibinfo {author} {\bibfnamefont {Juan}\ \bibnamefont
  {Carrasquilla}}, \bibinfo {author} {\bibfnamefont {Matthias}\ \bibnamefont
  {Troyer}}, \bibinfo {author} {\bibfnamefont {Roger}\ \bibnamefont {Melko}}, \
  and\ \bibinfo {author} {\bibfnamefont {Giuseppe}\ \bibnamefont {Carleo}},\
  }\bibfield  {title} {\enquote {\bibinfo {title} {Neural-network quantum state
  tomography},}\ }\href@noop {} {\bibfield  {journal} {\bibinfo  {journal}
  {Nat. Phys.}\ }\textbf {\bibinfo {volume} {14}},\ \bibinfo {pages} {447}
  (\bibinfo {year} {2018})}\BibitemShut {NoStop}%
\bibitem [{\citenamefont {Palmieri}\ \emph {et~al.}(2020)\citenamefont
  {Palmieri}, \citenamefont {Kovlakov}, \citenamefont {Bianchi}, \citenamefont
  {Yudin}, \citenamefont {Straupe}, \citenamefont {Biamonte},\ and\
  \citenamefont {Kulik}}]{palmieri2020experimental}%
  \BibitemOpen
  \bibfield  {author} {\bibinfo {author} {\bibfnamefont {Adriano~Macarone}\
  \bibnamefont {Palmieri}}, \bibinfo {author} {\bibfnamefont {Egor}\
  \bibnamefont {Kovlakov}}, \bibinfo {author} {\bibfnamefont {Federico}\
  \bibnamefont {Bianchi}}, \bibinfo {author} {\bibfnamefont {Dmitry}\
  \bibnamefont {Yudin}}, \bibinfo {author} {\bibfnamefont {Stanislav}\
  \bibnamefont {Straupe}}, \bibinfo {author} {\bibfnamefont {Jacob~D}\
  \bibnamefont {Biamonte}}, \ and\ \bibinfo {author} {\bibfnamefont {Sergei}\
  \bibnamefont {Kulik}},\ }\bibfield  {title} {\enquote {\bibinfo {title}
  {Experimental neural network enhanced quantum tomography},}\ }\href@noop {}
  {\bibfield  {journal} {\bibinfo  {journal} {npj Quantum Inf.}\ }\textbf
  {\bibinfo {volume} {6}},\ \bibinfo {pages} {1--5} (\bibinfo {year}
  {2020})}\BibitemShut {NoStop}%
\bibitem [{\citenamefont {Quek}\ \emph {et~al.}(2018)\citenamefont {Quek},
  \citenamefont {Fort},\ and\ \citenamefont {Ng}}]{quek2018adaptive}%
  \BibitemOpen
  \bibfield  {author} {\bibinfo {author} {\bibfnamefont {Yihui}\ \bibnamefont
  {Quek}}, \bibinfo {author} {\bibfnamefont {Stanislav}\ \bibnamefont {Fort}},
  \ and\ \bibinfo {author} {\bibfnamefont {Hui~Khoon}\ \bibnamefont {Ng}},\
  }\bibfield  {title} {\enquote {\bibinfo {title} {Adaptive quantum state
  tomography with neural networks},}\ }\href@noop {} {\bibfield  {journal}
  {\bibinfo  {journal} {arXiv:1812.06693}\ } (\bibinfo {year}
  {2018})}\BibitemShut {NoStop}%
\bibitem [{\citenamefont {Ahmed}\ \emph
  {et~al.}(2020{\natexlab{a}})\citenamefont {Ahmed}, \citenamefont {Mu{\~n}oz},
  \citenamefont {Nori},\ and\ \citenamefont
  {Kockum}}]{ahmed2020classification}%
  \BibitemOpen
  \bibfield  {author} {\bibinfo {author} {\bibfnamefont {Shahnawaz}\
  \bibnamefont {Ahmed}}, \bibinfo {author} {\bibfnamefont {Carlos~S{\'a}nchez}\
  \bibnamefont {Mu{\~n}oz}}, \bibinfo {author} {\bibfnamefont {Franco}\
  \bibnamefont {Nori}}, \ and\ \bibinfo {author} {\bibfnamefont {Anton~Frisk}\
  \bibnamefont {Kockum}},\ }\bibfield  {title} {\enquote {\bibinfo {title}
  {Classification and reconstruction of optical quantum states with deep neural
  networks},}\ }\href@noop {} {\bibfield  {journal} {\bibinfo  {journal}
  {arXiv:2012.02185}\ } (\bibinfo {year} {2020}{\natexlab{a}})}\BibitemShut
  {NoStop}%
\bibitem [{\citenamefont {Carrasquilla}\ \emph {et~al.}(2019)\citenamefont
  {Carrasquilla}, \citenamefont {Torlai}, \citenamefont {Melko},\ and\
  \citenamefont {Aolita}}]{carrasquilla2019reconstructing}%
  \BibitemOpen
  \bibfield  {author} {\bibinfo {author} {\bibfnamefont {Juan}\ \bibnamefont
  {Carrasquilla}}, \bibinfo {author} {\bibfnamefont {Giacomo}\ \bibnamefont
  {Torlai}}, \bibinfo {author} {\bibfnamefont {Roger~G}\ \bibnamefont {Melko}},
  \ and\ \bibinfo {author} {\bibfnamefont {Leandro}\ \bibnamefont {Aolita}},\
  }\bibfield  {title} {\enquote {\bibinfo {title} {Reconstructing quantum
  states with generative models},}\ }\href@noop {} {\bibfield  {journal}
  {\bibinfo  {journal} {Nat. Mach. Intell}\ }\textbf {\bibinfo {volume} {1}},\
  \bibinfo {pages} {155--161} (\bibinfo {year} {2019})}\BibitemShut {NoStop}%
\bibitem [{\citenamefont {Ahmed}\ \emph
  {et~al.}(2020{\natexlab{b}})\citenamefont {Ahmed}, \citenamefont {Mu{\~n}oz},
  \citenamefont {Nori},\ and\ \citenamefont {Kockum}}]{ahmed2020quantum}%
  \BibitemOpen
  \bibfield  {author} {\bibinfo {author} {\bibfnamefont {Shahnawaz}\
  \bibnamefont {Ahmed}}, \bibinfo {author} {\bibfnamefont {Carlos~S{\'a}nchez}\
  \bibnamefont {Mu{\~n}oz}}, \bibinfo {author} {\bibfnamefont {Franco}\
  \bibnamefont {Nori}}, \ and\ \bibinfo {author} {\bibfnamefont {Anton~Frisk}\
  \bibnamefont {Kockum}},\ }\bibfield  {title} {\enquote {\bibinfo {title}
  {Quantum state tomography with conditional generative adversarial
  networks},}\ }\href@noop {} {\bibfield  {journal} {\bibinfo  {journal}
  {arXiv:2008.03240}\ } (\bibinfo {year} {2020}{\natexlab{b}})}\BibitemShut
  {NoStop}%
\bibitem [{\citenamefont {Wetzstein}\ \emph {et~al.}(2020)\citenamefont
  {Wetzstein}, \citenamefont {Ozcan}, \citenamefont {Gigan}, \citenamefont
  {Fan}, \citenamefont {Englund}, \citenamefont {Solja{\v{c}}i{\'c}},
  \citenamefont {Denz}, \citenamefont {Miller},\ and\ \citenamefont
  {Psaltis}}]{wetzstein2020inference}%
  \BibitemOpen
  \bibfield  {author} {\bibinfo {author} {\bibfnamefont {Gordon}\ \bibnamefont
  {Wetzstein}}, \bibinfo {author} {\bibfnamefont {Aydogan}\ \bibnamefont
  {Ozcan}}, \bibinfo {author} {\bibfnamefont {Sylvain}\ \bibnamefont {Gigan}},
  \bibinfo {author} {\bibfnamefont {Shanhui}\ \bibnamefont {Fan}}, \bibinfo
  {author} {\bibfnamefont {Dirk}\ \bibnamefont {Englund}}, \bibinfo {author}
  {\bibfnamefont {Marin}\ \bibnamefont {Solja{\v{c}}i{\'c}}}, \bibinfo {author}
  {\bibfnamefont {Cornelia}\ \bibnamefont {Denz}}, \bibinfo {author}
  {\bibfnamefont {David~AB}\ \bibnamefont {Miller}}, \ and\ \bibinfo {author}
  {\bibfnamefont {Demetri}\ \bibnamefont {Psaltis}},\ }\bibfield  {title}
  {\enquote {\bibinfo {title} {Inference in artificial intelligence with deep
  optics and photonics},}\ }\href@noop {} {\bibfield  {journal} {\bibinfo
  {journal} {Nature}\ }\textbf {\bibinfo {volume} {588}},\ \bibinfo {pages}
  {39--47} (\bibinfo {year} {2020})}\BibitemShut {NoStop}%
\bibitem [{\citenamefont {Shastri}\ \emph {et~al.}(2021)\citenamefont
  {Shastri}, \citenamefont {Tait}, \citenamefont {de~Lima}, \citenamefont
  {Pernice}, \citenamefont {Bhaskaran}, \citenamefont {Wright},\ and\
  \citenamefont {Prucnal}}]{shastri2021photonics}%
  \BibitemOpen
  \bibfield  {author} {\bibinfo {author} {\bibfnamefont {Bhavin~J}\
  \bibnamefont {Shastri}}, \bibinfo {author} {\bibfnamefont {Alexander~N}\
  \bibnamefont {Tait}}, \bibinfo {author} {\bibfnamefont {T~Ferreira}\
  \bibnamefont {de~Lima}}, \bibinfo {author} {\bibfnamefont {Wolfram~HP}\
  \bibnamefont {Pernice}}, \bibinfo {author} {\bibfnamefont {Harish}\
  \bibnamefont {Bhaskaran}}, \bibinfo {author} {\bibfnamefont {C~David}\
  \bibnamefont {Wright}}, \ and\ \bibinfo {author} {\bibfnamefont {Paul~R}\
  \bibnamefont {Prucnal}},\ }\bibfield  {title} {\enquote {\bibinfo {title}
  {Photonics for artificial intelligence and neuromorphic computing},}\
  }\href@noop {} {\bibfield  {journal} {\bibinfo  {journal} {Nat. Photonics}\
  }\textbf {\bibinfo {volume} {15}},\ \bibinfo {pages} {102--114} (\bibinfo
  {year} {2021})}\BibitemShut {NoStop}%
\bibitem [{\citenamefont {Woods}\ and\ \citenamefont
  {Naughton}(2012)}]{woods2012photonic}%
  \BibitemOpen
  \bibfield  {author} {\bibinfo {author} {\bibfnamefont {Damien}\ \bibnamefont
  {Woods}}\ and\ \bibinfo {author} {\bibfnamefont {Thomas~J}\ \bibnamefont
  {Naughton}},\ }\bibfield  {title} {\enquote {\bibinfo {title} {Photonic
  neural networks},}\ }\href@noop {} {\bibfield  {journal} {\bibinfo  {journal}
  {Nat. Phys.}\ }\textbf {\bibinfo {volume} {8}},\ \bibinfo {pages} {257--259}
  (\bibinfo {year} {2012})}\BibitemShut {NoStop}%
\bibitem [{\citenamefont {Shen}\ \emph
  {et~al.}(2017{\natexlab{a}})\citenamefont {Shen}, \citenamefont {Harris},
  \citenamefont {Skirlo}, \citenamefont {Prabhu}, \citenamefont {Baehr-Jones},
  \citenamefont {Hochberg}, \citenamefont {Sun}, \citenamefont {Zhao},
  \citenamefont {Larochelle}, \citenamefont {Englund} \emph
  {et~al.}}]{shen2017deep}%
  \BibitemOpen
  \bibfield  {author} {\bibinfo {author} {\bibfnamefont {Yichen}\ \bibnamefont
  {Shen}}, \bibinfo {author} {\bibfnamefont {Nicholas~C}\ \bibnamefont
  {Harris}}, \bibinfo {author} {\bibfnamefont {Scott}\ \bibnamefont {Skirlo}},
  \bibinfo {author} {\bibfnamefont {Mihika}\ \bibnamefont {Prabhu}}, \bibinfo
  {author} {\bibfnamefont {Tom}\ \bibnamefont {Baehr-Jones}}, \bibinfo {author}
  {\bibfnamefont {Michael}\ \bibnamefont {Hochberg}}, \bibinfo {author}
  {\bibfnamefont {Xin}\ \bibnamefont {Sun}}, \bibinfo {author} {\bibfnamefont
  {Shijie}\ \bibnamefont {Zhao}}, \bibinfo {author} {\bibfnamefont {Hugo}\
  \bibnamefont {Larochelle}}, \bibinfo {author} {\bibfnamefont {Dirk}\
  \bibnamefont {Englund}},  \emph {et~al.},\ }\bibfield  {title} {\enquote
  {\bibinfo {title} {Deep learning with coherent nanophotonic circuits},}\
  }\href@noop {} {\bibfield  {journal} {\bibinfo  {journal} {Nat. Photonics}\
  }\textbf {\bibinfo {volume} {11}},\ \bibinfo {pages} {441} (\bibinfo {year}
  {2017}{\natexlab{a}})}\BibitemShut {NoStop}%
\bibitem [{\citenamefont {Lin}\ \emph {et~al.}(2018)\citenamefont {Lin},
  \citenamefont {Rivenson}, \citenamefont {Yardimci}, \citenamefont {Veli},
  \citenamefont {Luo}, \citenamefont {Jarrahi},\ and\ \citenamefont
  {Ozcan}}]{lin2018all}%
  \BibitemOpen
  \bibfield  {author} {\bibinfo {author} {\bibfnamefont {Xing}\ \bibnamefont
  {Lin}}, \bibinfo {author} {\bibfnamefont {Yair}\ \bibnamefont {Rivenson}},
  \bibinfo {author} {\bibfnamefont {Nezih~T}\ \bibnamefont {Yardimci}},
  \bibinfo {author} {\bibfnamefont {Muhammed}\ \bibnamefont {Veli}}, \bibinfo
  {author} {\bibfnamefont {Yi}~\bibnamefont {Luo}}, \bibinfo {author}
  {\bibfnamefont {Mona}\ \bibnamefont {Jarrahi}}, \ and\ \bibinfo {author}
  {\bibfnamefont {Aydogan}\ \bibnamefont {Ozcan}},\ }\bibfield  {title}
  {\enquote {\bibinfo {title} {All-optical machine learning using diffractive
  deep neural networks},}\ }\href@noop {} {\bibfield  {journal} {\bibinfo
  {journal} {Science}\ }\textbf {\bibinfo {volume} {361}},\ \bibinfo {pages}
  {1004--1008} (\bibinfo {year} {2018})}\BibitemShut {NoStop}%
\bibitem [{\citenamefont {Zuo}\ \emph {et~al.}(2019)\citenamefont {Zuo},
  \citenamefont {Li}, \citenamefont {Zhao}, \citenamefont {Jiang},
  \citenamefont {Chen}, \citenamefont {Chen}, \citenamefont {Jo}, \citenamefont
  {Liu},\ and\ \citenamefont {Du}}]{zuo2019all}%
  \BibitemOpen
  \bibfield  {author} {\bibinfo {author} {\bibfnamefont {Ying}\ \bibnamefont
  {Zuo}}, \bibinfo {author} {\bibfnamefont {Bohan}\ \bibnamefont {Li}},
  \bibinfo {author} {\bibfnamefont {Yujun}\ \bibnamefont {Zhao}}, \bibinfo
  {author} {\bibfnamefont {Yue}\ \bibnamefont {Jiang}}, \bibinfo {author}
  {\bibfnamefont {You-Chiuan}\ \bibnamefont {Chen}}, \bibinfo {author}
  {\bibfnamefont {Peng}\ \bibnamefont {Chen}}, \bibinfo {author} {\bibfnamefont
  {Gyu-Boong}\ \bibnamefont {Jo}}, \bibinfo {author} {\bibfnamefont {Junwei}\
  \bibnamefont {Liu}}, \ and\ \bibinfo {author} {\bibfnamefont {Shengwang}\
  \bibnamefont {Du}},\ }\bibfield  {title} {\enquote {\bibinfo {title}
  {All-optical neural network with nonlinear activation functions},}\
  }\href@noop {} {\bibfield  {journal} {\bibinfo  {journal} {Optica}\ }\textbf
  {\bibinfo {volume} {6}},\ \bibinfo {pages} {1132--1137} (\bibinfo {year}
  {2019})}\BibitemShut {NoStop}%
\bibitem [{\citenamefont {Zuo}\ \emph {et~al.}(2021)\citenamefont {Zuo},
  \citenamefont {Zhao}, \citenamefont {Chen}, \citenamefont {Du},\ and\
  \citenamefont {Liu}}]{zuo2021scalability}%
  \BibitemOpen
  \bibfield  {author} {\bibinfo {author} {\bibfnamefont {Ying}\ \bibnamefont
  {Zuo}}, \bibinfo {author} {\bibfnamefont {Yujun}\ \bibnamefont {Zhao}},
  \bibinfo {author} {\bibfnamefont {You-Chiuan}\ \bibnamefont {Chen}}, \bibinfo
  {author} {\bibfnamefont {Shengwang}\ \bibnamefont {Du}}, \ and\ \bibinfo
  {author} {\bibfnamefont {Junwei}\ \bibnamefont {Liu}},\ }\bibfield  {title}
  {\enquote {\bibinfo {title} {Scalability of all-optical neural networks based
  on spatial light modulators},}\ }\href {\doibase
  10.1103/PhysRevApplied.15.054034} {\bibfield  {journal} {\bibinfo  {journal}
  {Phys. Rev. Applied}\ }\textbf {\bibinfo {volume} {15}},\ \bibinfo {pages}
  {054034} (\bibinfo {year} {2021})}\BibitemShut {NoStop}%
\bibitem [{\citenamefont {Feldmann}\ \emph {et~al.}(2019)\citenamefont
  {Feldmann}, \citenamefont {Youngblood}, \citenamefont {Wright}, \citenamefont
  {Bhaskaran},\ and\ \citenamefont {Pernice}}]{feldmann2019all}%
  \BibitemOpen
  \bibfield  {author} {\bibinfo {author} {\bibfnamefont {J}~\bibnamefont
  {Feldmann}}, \bibinfo {author} {\bibfnamefont {N}~\bibnamefont {Youngblood}},
  \bibinfo {author} {\bibfnamefont {C~David}\ \bibnamefont {Wright}}, \bibinfo
  {author} {\bibfnamefont {H}~\bibnamefont {Bhaskaran}}, \ and\ \bibinfo
  {author} {\bibfnamefont {WHP}\ \bibnamefont {Pernice}},\ }\bibfield  {title}
  {\enquote {\bibinfo {title} {All-optical spiking neurosynaptic networks with
  self-learning capabilities},}\ }\href@noop {} {\bibfield  {journal} {\bibinfo
   {journal} {Nature}\ }\textbf {\bibinfo {volume} {569}},\ \bibinfo {pages}
  {208--214} (\bibinfo {year} {2019})}\BibitemShut {NoStop}%
\bibitem [{\citenamefont {Guo}\ \emph {et~al.}(2021)\citenamefont {Guo},
  \citenamefont {Barrett}, \citenamefont {Wang},\ and\ \citenamefont
  {Lvovsky}}]{Guo:21}%
  \BibitemOpen
  \bibfield  {author} {\bibinfo {author} {\bibfnamefont {Xianxin}\ \bibnamefont
  {Guo}}, \bibinfo {author} {\bibfnamefont {Thomas~D.}\ \bibnamefont
  {Barrett}}, \bibinfo {author} {\bibfnamefont {Zhiming~M.}\ \bibnamefont
  {Wang}}, \ and\ \bibinfo {author} {\bibfnamefont {A.~I.}\ \bibnamefont
  {Lvovsky}},\ }\bibfield  {title} {\enquote {\bibinfo {title} {Backpropagation
  through nonlinear units for the all-optical training of neural networks},}\
  }\href {\doibase 10.1364/PRJ.411104} {\bibfield  {journal} {\bibinfo
  {journal} {Photon. Res.}\ }\textbf {\bibinfo {volume} {9}},\ \bibinfo {pages}
  {B71--B80} (\bibinfo {year} {2021})}\BibitemShut {NoStop}%
\bibitem [{\citenamefont {Ryou}\ \emph {et~al.}(2021)\citenamefont {Ryou},
  \citenamefont {Whitehead}, \citenamefont {Zhelyeznyakov}, \citenamefont
  {Anderson}, \citenamefont {Keskin}, \citenamefont {Bajcsy},\ and\
  \citenamefont {Majumdar}}]{ryou2021freespace}%
  \BibitemOpen
  \bibfield  {author} {\bibinfo {author} {\bibfnamefont {Albert}\ \bibnamefont
  {Ryou}}, \bibinfo {author} {\bibfnamefont {James}\ \bibnamefont {Whitehead}},
  \bibinfo {author} {\bibfnamefont {Maksym}\ \bibnamefont {Zhelyeznyakov}},
  \bibinfo {author} {\bibfnamefont {Paul}\ \bibnamefont {Anderson}}, \bibinfo
  {author} {\bibfnamefont {Cem}\ \bibnamefont {Keskin}}, \bibinfo {author}
  {\bibfnamefont {Michal}\ \bibnamefont {Bajcsy}}, \ and\ \bibinfo {author}
  {\bibfnamefont {Arka}\ \bibnamefont {Majumdar}},\ }\href@noop {} {\enquote
  {\bibinfo {title} {Free-space optical neural network based on thermal atomic
  nonlinearity},}\ } (\bibinfo {year} {2021}),\ \Eprint
  {http://arxiv.org/abs/2102.04464} {arXiv:2102.04464 [cs.ET]} \BibitemShut
  {NoStop}%
\bibitem [{\citenamefont {Gross}\ \emph {et~al.}(2010)\citenamefont {Gross},
  \citenamefont {Liu}, \citenamefont {Flammia}, \citenamefont {Becker},\ and\
  \citenamefont {Eisert}}]{gross2010quantum}%
  \BibitemOpen
  \bibfield  {author} {\bibinfo {author} {\bibfnamefont {David}\ \bibnamefont
  {Gross}}, \bibinfo {author} {\bibfnamefont {Yi-Kai}\ \bibnamefont {Liu}},
  \bibinfo {author} {\bibfnamefont {Steven~T}\ \bibnamefont {Flammia}},
  \bibinfo {author} {\bibfnamefont {Stephen}\ \bibnamefont {Becker}}, \ and\
  \bibinfo {author} {\bibfnamefont {Jens}\ \bibnamefont {Eisert}},\ }\bibfield
  {title} {\enquote {\bibinfo {title} {Quantum state tomography via compressed
  sensing},}\ }\href@noop {} {\bibfield  {journal} {\bibinfo  {journal} {Phys.
  Rev. Lett.}\ }\textbf {\bibinfo {volume} {105}},\ \bibinfo {pages} {150401}
  (\bibinfo {year} {2010})}\BibitemShut {NoStop}%
\bibitem [{\citenamefont {Flammia}\ \emph {et~al.}(2012)\citenamefont
  {Flammia}, \citenamefont {Gross}, \citenamefont {Liu},\ and\ \citenamefont
  {Eisert}}]{flammia2012quantum}%
  \BibitemOpen
  \bibfield  {author} {\bibinfo {author} {\bibfnamefont {Steven~T}\
  \bibnamefont {Flammia}}, \bibinfo {author} {\bibfnamefont {David}\
  \bibnamefont {Gross}}, \bibinfo {author} {\bibfnamefont {Yi-Kai}\
  \bibnamefont {Liu}}, \ and\ \bibinfo {author} {\bibfnamefont {Jens}\
  \bibnamefont {Eisert}},\ }\bibfield  {title} {\enquote {\bibinfo {title}
  {Quantum tomography via compressed sensing: error bounds, sample complexity
  and efficient estimators},}\ }\href@noop {} {\bibfield  {journal} {\bibinfo
  {journal} {New J. Phys.}\ }\textbf {\bibinfo {volume} {14}},\ \bibinfo
  {pages} {095022} (\bibinfo {year} {2012})}\BibitemShut {NoStop}%
\bibitem [{\citenamefont {Chen}\ \emph {et~al.}(2013)\citenamefont {Chen},
  \citenamefont {Dawkins}, \citenamefont {Ji}, \citenamefont {Johnston},
  \citenamefont {Kribs}, \citenamefont {Shultz},\ and\ \citenamefont
  {Zeng}}]{chen2013uniqueness}%
  \BibitemOpen
  \bibfield  {author} {\bibinfo {author} {\bibfnamefont {Jianxin}\ \bibnamefont
  {Chen}}, \bibinfo {author} {\bibfnamefont {Hillary}\ \bibnamefont {Dawkins}},
  \bibinfo {author} {\bibfnamefont {Zhengfeng}\ \bibnamefont {Ji}}, \bibinfo
  {author} {\bibfnamefont {Nathaniel}\ \bibnamefont {Johnston}}, \bibinfo
  {author} {\bibfnamefont {David}\ \bibnamefont {Kribs}}, \bibinfo {author}
  {\bibfnamefont {Frederic}\ \bibnamefont {Shultz}}, \ and\ \bibinfo {author}
  {\bibfnamefont {Bei}\ \bibnamefont {Zeng}},\ }\bibfield  {title} {\enquote
  {\bibinfo {title} {Uniqueness of quantum states compatible with given
  measurement results},}\ }\href@noop {} {\bibfield  {journal} {\bibinfo
  {journal} {Phys. Rev. A}\ }\textbf {\bibinfo {volume} {88}},\ \bibinfo
  {pages} {012109} (\bibinfo {year} {2013})}\BibitemShut {NoStop}%
\bibitem [{\citenamefont {Ma}\ \emph {et~al.}(2016)\citenamefont {Ma},
  \citenamefont {Jackson}, \citenamefont {Zhou}, \citenamefont {Chen},
  \citenamefont {Lu}, \citenamefont {Mazurek}, \citenamefont {Fisher},
  \citenamefont {Peng}, \citenamefont {Kribs}, \citenamefont {Resch} \emph
  {et~al.}}]{ma2016pure}%
  \BibitemOpen
  \bibfield  {author} {\bibinfo {author} {\bibfnamefont {Xian}\ \bibnamefont
  {Ma}}, \bibinfo {author} {\bibfnamefont {Tyler}\ \bibnamefont {Jackson}},
  \bibinfo {author} {\bibfnamefont {Hui}\ \bibnamefont {Zhou}}, \bibinfo
  {author} {\bibfnamefont {Jianxin}\ \bibnamefont {Chen}}, \bibinfo {author}
  {\bibfnamefont {Dawei}\ \bibnamefont {Lu}}, \bibinfo {author} {\bibfnamefont
  {Michael~D}\ \bibnamefont {Mazurek}}, \bibinfo {author} {\bibfnamefont
  {Kent~AG}\ \bibnamefont {Fisher}}, \bibinfo {author} {\bibfnamefont {Xinhua}\
  \bibnamefont {Peng}}, \bibinfo {author} {\bibfnamefont {David}\ \bibnamefont
  {Kribs}}, \bibinfo {author} {\bibfnamefont {Kevin~J}\ \bibnamefont {Resch}},
  \emph {et~al.},\ }\bibfield  {title} {\enquote {\bibinfo {title} {Pure-state
  tomography with the expectation value of pauli operators},}\ }\href@noop {}
  {\bibfield  {journal} {\bibinfo  {journal} {Phys. Rev. A}\ }\textbf {\bibinfo
  {volume} {93}},\ \bibinfo {pages} {032140} (\bibinfo {year}
  {2016})}\BibitemShut {NoStop}%
\bibitem [{\citenamefont {Metcalf}\ and\ \citenamefont {van~der
  Straten}(2003)}]{metcalf2003laser}%
  \BibitemOpen
  \bibfield  {author} {\bibinfo {author} {\bibfnamefont {Harold~J}\
  \bibnamefont {Metcalf}}\ and\ \bibinfo {author} {\bibfnamefont {Peter}\
  \bibnamefont {van~der Straten}},\ }\bibfield  {title} {\enquote {\bibinfo
  {title} {Laser cooling and trapping of atoms},}\ }\href@noop {} {\bibfield
  {journal} {\bibinfo  {journal} {J. Opt. Soc. Am. B}\ }\textbf {\bibinfo
  {volume} {20}},\ \bibinfo {pages} {887--908} (\bibinfo {year}
  {2003})}\BibitemShut {NoStop}%
\bibitem [{\citenamefont {Zhang}\ \emph {et~al.}(2012)\citenamefont {Zhang},
  \citenamefont {Chen}, \citenamefont {Liu}, \citenamefont {Zhou},
  \citenamefont {Loy}, \citenamefont {Wong},\ and\ \citenamefont
  {Du}}]{2DMOTRSI2012}%
  \BibitemOpen
  \bibfield  {author} {\bibinfo {author} {\bibfnamefont {Shanchao}\
  \bibnamefont {Zhang}}, \bibinfo {author} {\bibfnamefont {J.~F.}\ \bibnamefont
  {Chen}}, \bibinfo {author} {\bibfnamefont {Chang}\ \bibnamefont {Liu}},
  \bibinfo {author} {\bibfnamefont {Shuyu}\ \bibnamefont {Zhou}}, \bibinfo
  {author} {\bibfnamefont {M.~M.~T.}\ \bibnamefont {Loy}}, \bibinfo {author}
  {\bibfnamefont {G.~K.~L.}\ \bibnamefont {Wong}}, \ and\ \bibinfo {author}
  {\bibfnamefont {Shengwang}\ \bibnamefont {Du}},\ }\bibfield  {title}
  {\enquote {\bibinfo {title} {A dark-line two-dimensional magneto-optical trap
  of 85rb atoms with high optical depth},}\ }\href {\doibase 10.1063/1.4732818}
  {\bibfield  {journal} {\bibinfo  {journal} {Rev. Sci. Instrum.}\ }\textbf
  {\bibinfo {volume} {83}},\ \bibinfo {pages} {073102} (\bibinfo {year}
  {2012})}\BibitemShut {NoStop}%
\bibitem [{\citenamefont {Harris}(1997)}]{EITHarris}%
  \BibitemOpen
  \bibfield  {author} {\bibinfo {author} {\bibfnamefont {Stephen~E}\
  \bibnamefont {Harris}},\ }\bibfield  {title} {\enquote {\bibinfo {title}
  {Electromagnetically induced transparency},}\ }\href@noop {} {\bibfield
  {journal} {\bibinfo  {journal} {Phys. Today}\ }\textbf {\bibinfo {volume}
  {50(7)}},\ \bibinfo {pages} {36--42} (\bibinfo {year} {1997})}\BibitemShut
  {NoStop}%
\bibitem [{\citenamefont {Fleischhauer}\ \emph {et~al.}(2005)\citenamefont
  {Fleischhauer}, \citenamefont {Imamoglu},\ and\ \citenamefont
  {Marangos}}]{fleischhauer2005electromagnetically}%
  \BibitemOpen
  \bibfield  {author} {\bibinfo {author} {\bibfnamefont {Michael}\ \bibnamefont
  {Fleischhauer}}, \bibinfo {author} {\bibfnamefont {Atac}\ \bibnamefont
  {Imamoglu}}, \ and\ \bibinfo {author} {\bibfnamefont {Jonathan~P}\
  \bibnamefont {Marangos}},\ }\bibfield  {title} {\enquote {\bibinfo {title}
  {Electromagnetically induced transparency: Optics in coherent media},}\
  }\href@noop {} {\bibfield  {journal} {\bibinfo  {journal} {Rev. Mod. Phys.}\
  }\textbf {\bibinfo {volume} {77}},\ \bibinfo {pages} {633} (\bibinfo {year}
  {2005})}\BibitemShut {NoStop}%
\bibitem [{ibm(2020)}]{ibmq_5_ourense}%
  \BibitemOpen
  \href@noop {} {\bibfield  {journal} {\bibinfo  {journal} {5-qubit backend:
  IBM Q team, ``IBM Q 5 Ourense backend specification V1.3.5,". Retrieved from
  https://quantum-computing.ibm.com}\ } (\bibinfo {year} {2020})}\BibitemShut
  {NoStop}%
\bibitem [{\citenamefont {Shen}\ \emph
  {et~al.}(2017{\natexlab{b}})\citenamefont {Shen}, \citenamefont {Harris},
  \citenamefont {Skirlo}, \citenamefont {Prabhu}, \citenamefont {Baehr-Jones},
  \citenamefont {Hochberg}, \citenamefont {Sun}, \citenamefont {Zhao},
  \citenamefont {Larochelle}, \citenamefont {Englund},\ and\ \citenamefont
  {Soljačić}}]{ShenNP2017}%
  \BibitemOpen
  \bibfield  {author} {\bibinfo {author} {\bibfnamefont {Yichen}\ \bibnamefont
  {Shen}}, \bibinfo {author} {\bibfnamefont {Nicholas~C.}\ \bibnamefont
  {Harris}}, \bibinfo {author} {\bibfnamefont {Scott}\ \bibnamefont {Skirlo}},
  \bibinfo {author} {\bibfnamefont {Mihika}\ \bibnamefont {Prabhu}}, \bibinfo
  {author} {\bibfnamefont {Tom}\ \bibnamefont {Baehr-Jones}}, \bibinfo {author}
  {\bibfnamefont {Michael}\ \bibnamefont {Hochberg}}, \bibinfo {author}
  {\bibfnamefont {Xin}\ \bibnamefont {Sun}}, \bibinfo {author} {\bibfnamefont
  {Shijie}\ \bibnamefont {Zhao}}, \bibinfo {author} {\bibfnamefont {Hugo}\
  \bibnamefont {Larochelle}}, \bibinfo {author} {\bibfnamefont {Dirk}\
  \bibnamefont {Englund}}, \ and\ \bibinfo {author} {\bibfnamefont {Marin}\
  \bibnamefont {Soljačić}},\ }\bibfield  {title} {\enquote {\bibinfo {title}
  {Deep learning with coherent nanophotonic circuits},}\ }\href@noop {}
  {\bibfield  {journal} {\bibinfo  {journal} {Nat. Photonics}\ }\textbf
  {\bibinfo {volume} {11}},\ \bibinfo {pages} {441--446} (\bibinfo {year}
  {2017}{\natexlab{b}})}\BibitemShut {NoStop}%
\bibitem [{\citenamefont {Kimble}(2008)}]{Kimble2008}%
  \BibitemOpen
  \bibfield  {author} {\bibinfo {author} {\bibfnamefont {H.~J.}\ \bibnamefont
  {Kimble}},\ }\bibfield  {title} {\enquote {\bibinfo {title} {{The quantum
  internet}},}\ }\href {\doibase 10.1038/nature07127} {\bibfield  {journal}
  {\bibinfo  {journal} {Nature}\ }\textbf {\bibinfo {volume} {453}},\ \bibinfo
  {pages} {1023--1030} (\bibinfo {year} {2008})},\ \Eprint
  {http://arxiv.org/abs/0806.4195} {0806.4195} \BibitemShut {NoStop}%
\bibitem [{\citenamefont {Gu}\ \emph {et~al.}(2019)\citenamefont {Gu},
  \citenamefont {Yang},\ and\ \citenamefont {Chen}}]{Gu2019}%
  \BibitemOpen
  \bibfield  {author} {\bibinfo {author} {\bibfnamefont {Zhenjie}\ \bibnamefont
  {Gu}}, \bibinfo {author} {\bibfnamefont {Ce}~\bibnamefont {Yang}}, \ and\
  \bibinfo {author} {\bibfnamefont {J.~F.}\ \bibnamefont {Chen}},\ }\bibfield
  {title} {\enquote {\bibinfo {title} {{Characterization of the photon-number
  state of a narrow-band single photon generated from a cold atomic cloud}},}\
  }\href {\doibase 10.1016/j.optcom.2019.01.074} {\bibfield  {journal}
  {\bibinfo  {journal} {Opt. Commun.}\ }\textbf {\bibinfo {volume} {439}},\
  \bibinfo {pages} {206--209} (\bibinfo {year} {2019})}\BibitemShut {NoStop}%
\bibitem [{\citenamefont {Yu}\ \emph {et~al.}(2020)\citenamefont {Yu},
  \citenamefont {Ma}, \citenamefont {Luo}, \citenamefont {Jing}, \citenamefont
  {Sun}, \citenamefont {Fang}, \citenamefont {Yang}, \citenamefont {Liu},
  \citenamefont {Zheng}, \citenamefont {Xie}, \citenamefont {Zhang},
  \citenamefont {You}, \citenamefont {Wang}, \citenamefont {Chen},
  \citenamefont {Zhang}, \citenamefont {Bao},\ and\ \citenamefont
  {Pan}}]{Yu2020}%
  \BibitemOpen
  \bibfield  {author} {\bibinfo {author} {\bibfnamefont {Yong}\ \bibnamefont
  {Yu}}, \bibinfo {author} {\bibfnamefont {Fei}\ \bibnamefont {Ma}}, \bibinfo
  {author} {\bibfnamefont {Xi~Yu}\ \bibnamefont {Luo}}, \bibinfo {author}
  {\bibfnamefont {Bo}~\bibnamefont {Jing}}, \bibinfo {author} {\bibfnamefont
  {Peng~Fei}\ \bibnamefont {Sun}}, \bibinfo {author} {\bibfnamefont {Ren~Zhou}\
  \bibnamefont {Fang}}, \bibinfo {author} {\bibfnamefont {Chao~Wei}\
  \bibnamefont {Yang}}, \bibinfo {author} {\bibfnamefont {Hui}\ \bibnamefont
  {Liu}}, \bibinfo {author} {\bibfnamefont {Ming~Yang}\ \bibnamefont {Zheng}},
  \bibinfo {author} {\bibfnamefont {Xiu~Ping}\ \bibnamefont {Xie}}, \bibinfo
  {author} {\bibfnamefont {Wei~Jun}\ \bibnamefont {Zhang}}, \bibinfo {author}
  {\bibfnamefont {Li~Xing}\ \bibnamefont {You}}, \bibinfo {author}
  {\bibfnamefont {Zhen}\ \bibnamefont {Wang}}, \bibinfo {author} {\bibfnamefont
  {Teng~Yun}\ \bibnamefont {Chen}}, \bibinfo {author} {\bibfnamefont {Qiang}\
  \bibnamefont {Zhang}}, \bibinfo {author} {\bibfnamefont {Xiao~Hui}\
  \bibnamefont {Bao}}, \ and\ \bibinfo {author} {\bibfnamefont {Jian~Wei}\
  \bibnamefont {Pan}},\ }\bibfield  {title} {\enquote {\bibinfo {title}
  {{Entanglement of two quantum memories via fibres over dozens of
  kilometres}},}\ }\href {\doibase 10.1038/s41586-020-1976-7} {\bibfield
  {journal} {\bibinfo  {journal} {Nature}\ }\textbf {\bibinfo {volume} {578}},\
  \bibinfo {pages} {240--245} (\bibinfo {year} {2020})},\ \Eprint
  {http://arxiv.org/abs/1903.11284} {1903.11284} \BibitemShut {NoStop}%
\bibitem [{\citenamefont {Li}\ \emph {et~al.}(2020)\citenamefont {Li},
  \citenamefont {Jiang}, \citenamefont {Wu}, \citenamefont {Chang},
  \citenamefont {Pu}, \citenamefont {Zhang},\ and\ \citenamefont
  {Duan}}]{Li2020}%
  \BibitemOpen
  \bibfield  {author} {\bibinfo {author} {\bibfnamefont {C.}~\bibnamefont
  {Li}}, \bibinfo {author} {\bibfnamefont {N.}~\bibnamefont {Jiang}}, \bibinfo
  {author} {\bibfnamefont {Y.~K.}\ \bibnamefont {Wu}}, \bibinfo {author}
  {\bibfnamefont {W.}~\bibnamefont {Chang}}, \bibinfo {author} {\bibfnamefont
  {Y.~F.}\ \bibnamefont {Pu}}, \bibinfo {author} {\bibfnamefont
  {S.}~\bibnamefont {Zhang}}, \ and\ \bibinfo {author} {\bibfnamefont {L.~M.}\
  \bibnamefont {Duan}},\ }\bibfield  {title} {\enquote {\bibinfo {title}
  {{Quantum Communication between Multiplexed Atomic Quantum Memories}},}\
  }\href {\doibase 10.1103/PhysRevLett.124.240504} {\bibfield  {journal}
  {\bibinfo  {journal} {Phys. Rev. Lett.}\ }\textbf {\bibinfo {volume} {124}},\
  \bibinfo {pages} {1--6} (\bibinfo {year} {2020})},\ \Eprint
  {http://arxiv.org/abs/1909.02185} {1909.02185} \BibitemShut {NoStop}%
\bibitem [{\citenamefont {Choi}\ \emph {et~al.}(2010)\citenamefont {Choi},
  \citenamefont {Goban}, \citenamefont {Papp}, \citenamefont {{Van Enk}},\ and\
  \citenamefont {Kimble}}]{Choi2010}%
  \BibitemOpen
  \bibfield  {author} {\bibinfo {author} {\bibfnamefont {K.~S.}\ \bibnamefont
  {Choi}}, \bibinfo {author} {\bibfnamefont {A.}~\bibnamefont {Goban}},
  \bibinfo {author} {\bibfnamefont {S.~B.}\ \bibnamefont {Papp}}, \bibinfo
  {author} {\bibfnamefont {S.~J.}\ \bibnamefont {{Van Enk}}}, \ and\ \bibinfo
  {author} {\bibfnamefont {H.~J.}\ \bibnamefont {Kimble}},\ }\bibfield  {title}
  {\enquote {\bibinfo {title} {{Entanglement of spin waves among four quantum
  memories}},}\ }\href {\doibase 10.1038/nature09568} {\bibfield  {journal}
  {\bibinfo  {journal} {Nature}\ }\textbf {\bibinfo {volume} {468}},\ \bibinfo
  {pages} {412--418} (\bibinfo {year} {2010})},\ \Eprint
  {http://arxiv.org/abs/1007.1664} {1007.1664} \BibitemShut {NoStop}%
\bibitem [{\citenamefont {Pu}\ \emph {et~al.}(2018)\citenamefont {Pu},
  \citenamefont {Wu}, \citenamefont {Jiang}, \citenamefont {Chang},
  \citenamefont {Li}, \citenamefont {Zhang},\ and\ \citenamefont
  {Duan}}]{Pu2017}%
  \BibitemOpen
  \bibfield  {author} {\bibinfo {author} {\bibfnamefont {Yunfei}\ \bibnamefont
  {Pu}}, \bibinfo {author} {\bibfnamefont {Yukai}\ \bibnamefont {Wu}}, \bibinfo
  {author} {\bibfnamefont {Nan}\ \bibnamefont {Jiang}}, \bibinfo {author}
  {\bibfnamefont {Wei}\ \bibnamefont {Chang}}, \bibinfo {author} {\bibfnamefont
  {Chang}\ \bibnamefont {Li}}, \bibinfo {author} {\bibfnamefont {Sheng}\
  \bibnamefont {Zhang}}, \ and\ \bibinfo {author} {\bibfnamefont {Luming}\
  \bibnamefont {Duan}},\ }\bibfield  {title} {\enquote {\bibinfo {title}
  {Experimental entanglement of 25 individually accessible atomic quantum
  interfaces},}\ }\href@noop {} {\bibfield  {journal} {\bibinfo  {journal}
  {Sci. Adv.}\ }\textbf {\bibinfo {volume} {4}},\ \bibinfo {pages} {eaar3931}
  (\bibinfo {year} {2018})}\BibitemShut {NoStop}%
\bibitem [{\citenamefont {Xu}\ \emph {et~al.}(2015)\citenamefont {Xu},
  \citenamefont {Wang}, \citenamefont {Chen},\ and\ \citenamefont
  {Li}}]{xu2015empirical}%
  \BibitemOpen
  \bibfield  {author} {\bibinfo {author} {\bibfnamefont {Bing}\ \bibnamefont
  {Xu}}, \bibinfo {author} {\bibfnamefont {Naiyan}\ \bibnamefont {Wang}},
  \bibinfo {author} {\bibfnamefont {Tianqi}\ \bibnamefont {Chen}}, \ and\
  \bibinfo {author} {\bibfnamefont {Mu}~\bibnamefont {Li}},\ }\bibfield
  {title} {\enquote {\bibinfo {title} {Empirical evaluation of rectified
  activations in convolutional network},}\ }\href@noop {} {\bibfield  {journal}
  {\bibinfo  {journal} {arXiv:1505.00853}\ } (\bibinfo {year}
  {2015})}\BibitemShut {NoStop}%
\bibitem [{\citenamefont {Di~Leonardo}\ \emph {et~al.}(2007)\citenamefont
  {Di~Leonardo}, \citenamefont {Ianni},\ and\ \citenamefont
  {Ruocco}}]{di2007computer}%
  \BibitemOpen
  \bibfield  {author} {\bibinfo {author} {\bibfnamefont {Roberto}\ \bibnamefont
  {Di~Leonardo}}, \bibinfo {author} {\bibfnamefont {Francesca}\ \bibnamefont
  {Ianni}}, \ and\ \bibinfo {author} {\bibfnamefont {Giancarlo}\ \bibnamefont
  {Ruocco}},\ }\bibfield  {title} {\enquote {\bibinfo {title} {Computer
  generation of optimal holograms for optical trap arrays},}\ }\href@noop {}
  {\bibfield  {journal} {\bibinfo  {journal} {Opt. Express}\ }\textbf {\bibinfo
  {volume} {15}},\ \bibinfo {pages} {1913--1922} (\bibinfo {year}
  {2007})}\BibitemShut {NoStop}%
\bibitem [{\citenamefont {Nogrette}\ \emph {et~al.}(2014)\citenamefont
  {Nogrette}, \citenamefont {Labuhn}, \citenamefont {Ravets}, \citenamefont
  {Barredo}, \citenamefont {B{\'e}guin}, \citenamefont {Vernier}, \citenamefont
  {Lahaye},\ and\ \citenamefont {Browaeys}}]{nogrette2014single}%
  \BibitemOpen
  \bibfield  {author} {\bibinfo {author} {\bibfnamefont {Florence}\
  \bibnamefont {Nogrette}}, \bibinfo {author} {\bibfnamefont {Henning}\
  \bibnamefont {Labuhn}}, \bibinfo {author} {\bibfnamefont {Sylvain}\
  \bibnamefont {Ravets}}, \bibinfo {author} {\bibfnamefont {Daniel}\
  \bibnamefont {Barredo}}, \bibinfo {author} {\bibfnamefont {Lucas}\
  \bibnamefont {B{\'e}guin}}, \bibinfo {author} {\bibfnamefont {Aline}\
  \bibnamefont {Vernier}}, \bibinfo {author} {\bibfnamefont {Thierry}\
  \bibnamefont {Lahaye}}, \ and\ \bibinfo {author} {\bibfnamefont {Antoine}\
  \bibnamefont {Browaeys}},\ }\bibfield  {title} {\enquote {\bibinfo {title}
  {Single-atom trapping in holographic 2d arrays of microtraps with arbitrary
  geometries},}\ }\href@noop {} {\bibfield  {journal} {\bibinfo  {journal}
  {Phys. Rev. X}\ }\textbf {\bibinfo {volume} {4}},\ \bibinfo {pages} {021034}
  (\bibinfo {year} {2014})}\BibitemShut {NoStop}%
\end{thebibliography}%

\appendix

\section{Training and testing the NN}\label{app:ONN-sim}

The input layer is determined by the sampled Pauli operators, the number of neurons equals to $m$. The number of output layer neurons is equals to the independent amplitude parameters (see~\cref{eq:ket}), denote as $d$, e.g. $d$ equals $2\cdot2-1 = 3$ for 1-qubit,  $2\cdot2^2-1 = 7$ for 2-qubit, and $2\cdot2^3-1 = 15$ for 3-qubit.
There are four hidden layers for each network, each layer contains $32\cdot d$ neurons. The activation function between each layer is the popular Leaky ReLu~\cite{xu2015empirical}, a modified version of the classic non-linear activation function ReLu.
The network type is fully connected and feed-forward,
the optimizer is Adam, and the loss function is mean square error
\begin{equation}\label{eq:loss}
MSE(\phi) = E_{\phi}[(\hat{\phi} - \phi)^{2}],
\end{equation}
for all tested examples.

For each 2-qubit case, 20,000 pairs of training data are used for training. The corresponding number is 150,000 for each 3-qubit case. And the number of iterations is 300 for all networks.

\section{UDA operator sets}\label{app:UDA-op}

The UDA operator sets~\cite{ma2016pure} for 2- and 3-qubit are as follows:

\begin{align*}
P_\text{2-UDA} = \{&IX,IY,IZ,XI,YX,YY,\\
&Y Z,ZX,ZY,ZZ\},
\end{align*}
and
\begin{align*}
P_\text{3-UDA} = \{&IIX,IIY,IIZ,IXI,IXX,IXY,IYI,\\
&IYX,IYY,I ZI,XI Z,XXX,XXY,\\
&XYX,XYY,XZX,XZY,YXX,YXY,\\
&YXZ,YYX,YYY,YYZ,YZI,ZII,\\
&ZXZ,ZY Z,ZZX,ZZY,ZZZ\}.
\end{align*}

\section{Polarization analysis and Poincare sphere }\label{app:polar}

In describing polarizations, we use the convention of the $xyz$ coordinator system, i.e., we look toward the light source, then the z-axis is in the direction of light momentum. Thus the North Pole represents $\sigma_{+}$, i.e., left-rotated polarization and the South Pole represents $\sigma_{-}$, i.e., right-rotated polarization. $|H\rangle$ ($|\nearrow\rangle=\frac{1}{\sqrt{2}}(|H\rangle+|V\rangle)$) and $|V\rangle$ ($|\searrow\rangle=\frac{1}{\sqrt{2}}(|H\rangle-|V\rangle)$) locates at the intersections of $S_{1}$ ($S_{3}$) and the equator of $Poincar\acute{e}$ sphere. State $|\psi\rangle=\frac{1}{\sqrt{2}}(|H\rangle+e^{i\theta}|V\rangle)$ is on the intersection circle between $S_{2}S_{3}$ plane and the sphere, where $\theta$ is the angle of $|\psi\rangle$ to $|\nearrow\rangle$. 

\begin{figure} [t]
	\includegraphics[width=0.3\textwidth]{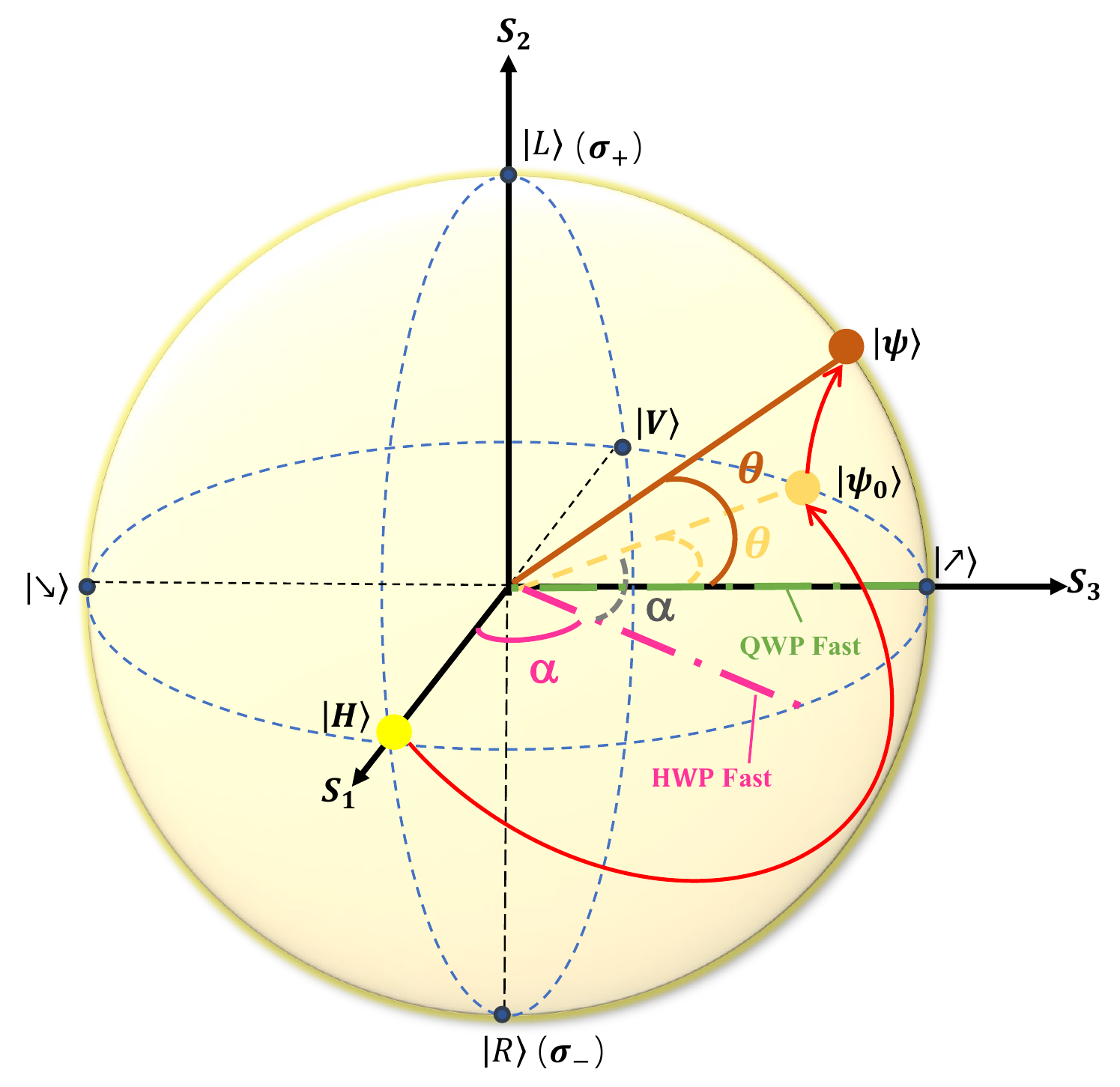}
	\caption{\label{fig:Poincare}(color onine) The $Poincar\acute{e}$ sphere. The initial state $|H\rangle$ can be rotated to $|\psi\rangle=\frac{1}{\sqrt{2}}(|H\rangle+e^{i\theta}|V\rangle)$ with the combination of a half-wave plate (HWP) and a quarter-wave plate (QWP). 
	} 
\end{figure}

 The initial state $|H\rangle$ is prepared with laser going through a polarization beam splitter (PBS). Then, we set the fast axis (pink line in Fig.~\ref{fig:Poincare} ) of HWP with an angle of $\alpha$ to $|H\rangle$, so $|H\rangle$ will rotate along the fast axis of HWP by $180^{\circ}$ to $|\psi_{0}\rangle$ (the light orange point in Figure 1), with the angle of $\theta$ to $S_{3}$. Finally, the fast axis of QWP (the green dot-line axis in Fig~\ref{fig:Poincare}) is set at the direction of $S_{3}$, so $|\psi_{0}\rangle$ will rotate along $S_{3}$ by $90^{\circ}$ to $|\psi\rangle$ (the dark orange point in Figure 1), with the angle to $S_{3}$ as $\theta$. Since the angle in the real world is the half of that in Poincare sphere, in experiment we rotate the fast axis of HWP with an angle of $\frac{\alpha}{2}$ to x-plus axis in the range from ${\frac{\pi}{8}}$ to ${\frac{\pi}{4}}$, and set the fast axis of QWP with an angle of $\frac{\pi}{4}$ to x-plus axis.

The measurement of $\langle Z \rangle$ is shown in III of Fig~\ref{fig:system}(a). The laser in polarization state $|\psi\rangle$ is incident on a PBS and the output powers of the two pannles are measured. Thus the probability of $|H\rangle$ is the transmittance T and the probability of $|V\rangle$ is the reflecitivity R. Since the eigenvalue of $|H\rangle$ and $|V\rangle$ are +1 and -1 respectively, $\langle Z \rangle=T-R$. As for the measurement results of $\langle X \rangle$ and $\langle Y \rangle$, we expect they have the same expressions as T-R, which means we should get both $|H\rangle$ if the inputs are $|\nearrow\rangle$ and $|L\rangle$ in these two measurements respectively. It is clear to see in $Poincar\acute{e}$ sphere that if we use a HWP with the fast axis having an angle of $\frac{\pi}{4}$ to $|H\rangle$ and a QWP with the fast axis at the direction of $S_{3}$ respectively, we can rotate the input state to get the expected measurement results. So II of Fig~\ref{fig:system}(a), we insert a HWP with an angle of $\frac{\pi}{8}$ to x-plus axis in front of the PBS to measure $\langle X \rangle$ and replace the HWP with a QWP with angle of $\frac{\pi}{4}$ to the x-plus axis to measure $\langle Y \rangle$ as shown in IV of Fig~\ref{fig:system}(a).

\section{Implementation of weighted beam generation}\label{app:gsw}
SLMs using in experiment conducted the input generation and two linear operations. For input generation, the first SLM (SLM1) is divided into rectangular parts with same size. Each part contributes to generate an element of input vector with range from 0 to 1. Without an aperture, two lenses consist of 4-f system and the exact same intensity distribution is imaged on the second SLM. We apply blazed grating and a sine modulation to the SLM with the form below:
\begin{equation}
\phi_{grating}(i,j)=\frac{2\pi}{T_i}i+\frac{2\pi}{T_j}j 
\end{equation}
\begin{equation}
\phi_{mod}(i,j)=m\pi \sin (\frac{2\pi}{T_{mi}}i+\frac{2\pi}{T_{mj}j})
\end{equation}
$\phi_{grating}(i,j) and \phi_{mod}(i,j)$ presents the grating phase and modulation phase applied on SLM.$i,j$ present the pixel index along two directions and $T_i$ and $T_j$ are the line spacing accordingly. $m$ is called modulation depth. $T_{mi}$ and $T_{mj}$ are the period of this sine function along two directions. On focal plane, the beam is separated into multiple beams due to modulation. The intensity of beam located on the same position of the beam with no modulation is determined by modulation depth $m$. Thus, the input can be modulated.

The linear operation conducted by SLM2 and SLM3. SLMs are divided to multiple parts and each part receive the output from the last layer and divided the output of last layer to multiple beams on the focal plane. We apply iterative algorithm to obtain optical holograms targeted to the generation of weighted multiple spots and weighted Gerchberg-Saxton (GSW) algorithm\cite{di2007computer} \cite{nogrette2014single} is one of frequently used algorithm. We applied several modifications on GSW algorithm to achieve better performance. Experimentally, the output beam intensity distribution is not as same as computer calculation. The difference can come from the deflections on optical path, SLM surface and inaccuracy of incident intensity. Theoretically, it’s hard to simulate all these deflections. We apply the adaptive iteration by using the experimentally measured intensity $I_n$ instead of numerical calculated amplitude to perform the iteration after FFT. i.e. We measure the intensity distribution by camera, if the difference between camera captured intensity and target intensity, we can drive this pattern to SLM. Otherwise, we replace the amplitude to $ g_n \sqrt{I_t}$ with $g_n =\frac{\sqrt{I_t }}{\sqrt{I_n}} g_{n-1}$ and do inverse FFT. To avoid over-compensation, we add parameter a to adjust the feedback as $g_n=a \frac{\sqrt{I_t}}{\sqrt{I_n}} g_{n-1}+(1-a)$, where $a$ is in the range $(0,1)$.

\section{Training Optical Neural Network}\label{app:AONN-train}
In the optical neural network tomography scheme, we input $\{1-\langle \sigma_{x} \rangle,1-\langle \sigma_{y} \rangle,1-\langle \sigma_{z} \rangle\}$ to the optical neural network instead of $\{\langle \sigma_{x} \rangle,\langle \sigma_{y} \rangle,\langle \sigma_{z} \rangle\}$  because the non-linear activation function 
\begin{equation}\label{eqn:EIT}
I_{p}^{out}=f(I_c)=I_{p}^{in}e^{-OD\frac{4\gamma_{12}\gamma_{13}}{\Omega_c^2+4\gamma_{12}\gamma_{13}} }
\end{equation}
is convex. We need to make sure the target regression function is convex as well.

\begin{figure}
	\includegraphics[width=0.5\textwidth]{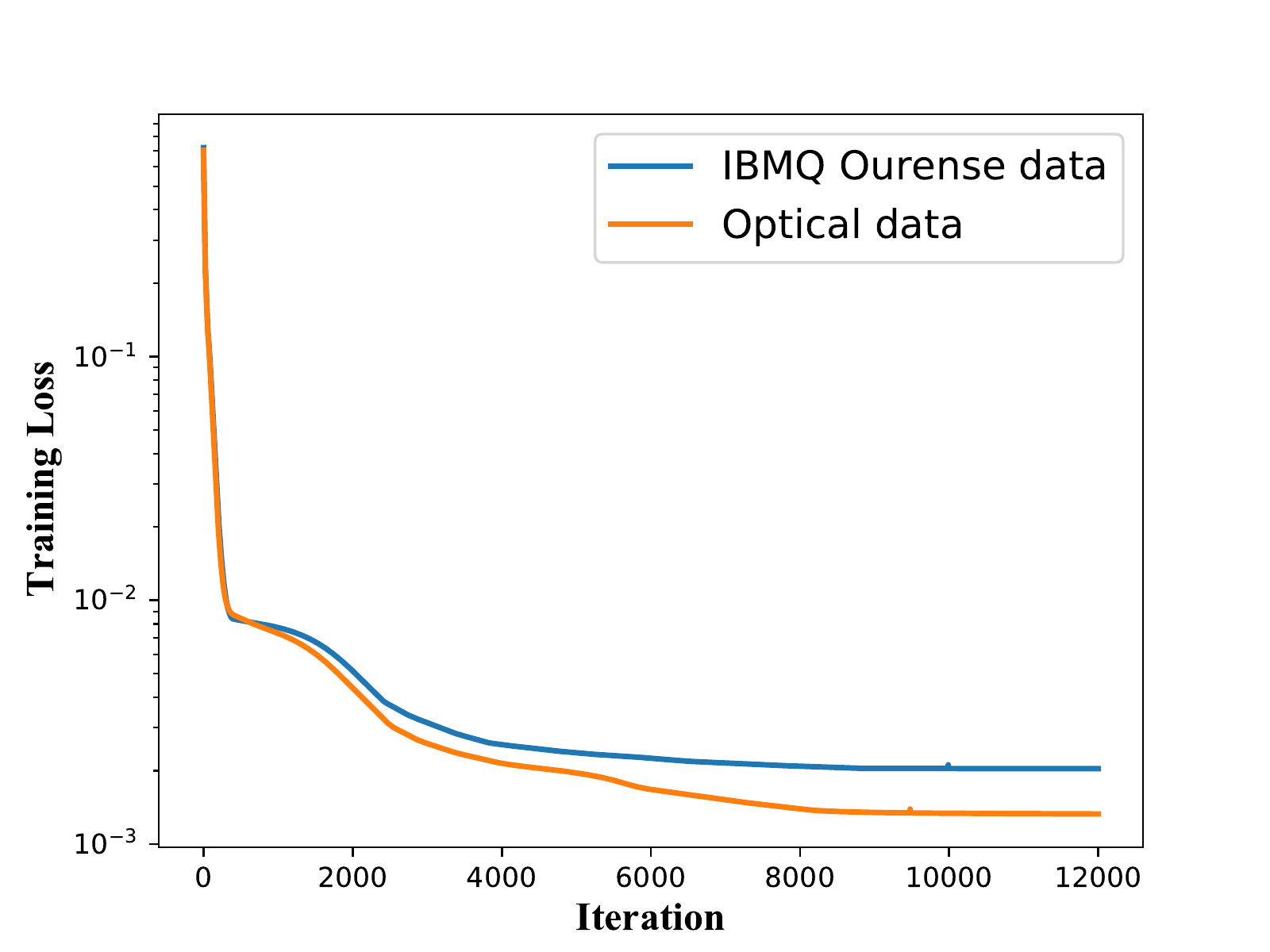}
	\caption{\label{fig:iter} Training loss versus the number of iterations.}
\end{figure}

In our experiments, the optical neural network is trained via classical simulation. The optimizer is Adam, the learning rate is 0.002, the loss function is also mean square error (\cref{eq:loss})
for both data sampled from our optical measurements and the IBM Ourense quantum computer, 10000 iterations suffice to converge, as shown in Fig. \ref{fig:iter}.

\end{document}